%% file: BrIcc_SH.tex
\documentclass[reviewcopy]{elsarticle}

\usepackage[reviewcopy]{adndt}
\usepackage{longtable}

\usepackage{amsmath}

\usepackage{amssymb}

\biboptions{square,sort&compress}
\bibpunct[]{[}{]}{,}{n}{}{;}
\def\phe{\phantom{0}}  
\def\phk{\phantom{00}} 
\begin{document}

\begin{frontmatter}

\journal{Atomic Data and Nuclear Data Tables}


\title{$\diamondsuit$ Conversion coefficients for superheavy elements}

  \author[ANU]{T. Kib\'edi\corref{cor1}}
  \ead{E-mail: Tibor.Kibedi@anu.edu.au}

  \author[Gatchina]{M.B. Trzhaskovskaya}

  \author[Manipal]{M. Gupta}

  \author[ANU]{A.E. Stuchbery}
  \cortext[cor1]{Corresponding author.}

  \address[ANU]{Department of Nuclear Physics, Research School of Physics and Engineering,
                The Australian National University, Canberra, ACT 0200, Australia}

  \address[Gatchina]{Petersburg Nuclear Physics Institute, Gatchina, Russia 188300}

  \address[Manipal]{Manipal University, Manipal 576104, Karnataka, India}

\date{02.11.2010} 

\begin{abstract}
 In this paper we report on internal conversion coefficients for $Z=111$ to $Z=126$ superheavy elements
 obtained from relativistic Dirac--Fock (DF) calculations.
 The effect of the atomic vacancy created during the conversion process has been taken into account
 using the so called \emph{``Frozen Orbital"} approximation.
 The selection of this atomic model is supported by our recent comparison of experimental and theoretical
 conversion coefficients across a wide range of nuclei.
 The atomic masses, valence shell electron configurations, and theoretical atomic binding energies
 required for the calculations were adopted from a critical evaluation of the published data.
 The new conversion coefficient data tables presented here cover all atomic shells,
 transition energies from $1$ keV up to $6000$ keV, and multipole orders of $1$ to $5$.
 A similar approach was used in our previous calculations \cite{int:2008Ki07} for $Z=5-110$. \\
 
 $\diamondsuit$ \emph{\emph{Accepted for publication in Atomic Data and Nuclear Data Tables}}
\end{abstract}

\end{frontmatter}




\newpage

\tableofcontents
\listofDtables
\listofDfigures
\vskip5pc

\include{Introduction}
\ack
The authors wish to thank V. Pershina for helpful discussions and materials.
Supported in part by the Australian Research Council Discovery Grant No. DP0773273.

\include{IntroReferences}
\include{TableExplanation}

\datatables            
\include{AtomicNuclearData}

\include{Z111_TotIcc}
\include{Z112_TotIcc}
\include{Z113_TotIcc}
\include{Z114_TotIcc}
\include{Z115_TotIcc}
\include{Z116_TotIcc}
\include{Z117_TotIcc}
\include{Z118_TotIcc}
\include{Z119_TotIcc}
\include{Z120_TotIcc}
\include{Z121_TotIcc}
\include{Z122_TotIcc}
\include{Z123_TotIcc}
\include{Z124_TotIcc}
\include{Z125_TotIcc}
\include{Z126_TotIcc}
\newpage

\include{TableReferences}

\end{document}

%% file: Introduction.tex
\section{Introduction}
\label{sec:intro}
 The superheavy elements (SHE) inhabit the nuclear landscape at the very top end of the nuclear
 chart.
 Existing only due to shell effects, the observation and characterization of their decay properties
 remains fundamentally important to understanding  the evolution of nuclear shape and stability at
 the very limits of mass and charge.
 The possibility of new regions of stability  beyond $Z=82$ and $N=126$ has been the driving force 
 for such research.
 Predictions have been made for heavier proton and neutron shell closures and are broadly based on 
 microscopic--macroscopic and self--consistent (relativistic and non--relativistic) nuclear 
 mean--field models.
 Most theories predict $N=184$ as being a closed shell whereas predictions for proton magic numbers
 vary, mainly due to differences in the treatment of the Coulomb term and the spin-orbit interaction.
 microscopic--macroscopic theories \cite{int:1997Mol,int:2003MunA,int:2003MunB} agree that 
 $^{298}$114$_{184}$ is doubly magic while non--relativistic microscopic models \cite{int:1999CwAA} 
 find that $Z=114$ and $120$ are equally likely shell--closures.
 Relativistic mean field theory suggests $N \sim 164$, $172$ and $184$ for $Z \sim118$ underscoring
 the fact that magic shells are strongly {\it isotope} dependent \cite{int:2005Gam}.
 Superheavy nuclides are very short lived and their synthesis presents a challenge to experimenters.
 While the primary decay mode is through $\alpha$ emission, spontaneous fission has also been observed.
 Another characteristic feature could be the high probability of conversion electron emission over
 photon emission due to the dependence of the electromagnetic processes on atomic number.
 An accurate knowledge of the theoretical internal conversion coefficients (i.e. the ratio of the
 conversion electrons to $\gamma$--rays), is important for both theoretical and experimental nuclear studies
  \cite{int:2008HeAA}.
 In this paper we report the calculation of theoretical internal conversion coefficients (ICC) for 
 $Z=111-126$ elements.
 The new tables together with the BrIcc \cite{int:2008Ki07} tables provide ICCs for all known 
 elements as required for nuclear research and applications.

\section{Calculation}
\label{sec:calculation}

 The emission of an atomic electron in a nuclear transition is a second order electrodynamical
 process, in which a virtual photon is exchanged between the nucleus and the electron.
 The ICC in the $i-$th atomic sub--shell, $\alpha_{i}$ is defined as the ratio of the number of
 electrons, $N_{i}$ to the number of gamma quanta, $N_{\gamma}$ ejected in the electromagnetic
 disintegration process:
 \begin{equation}
  \alpha_{i} = \frac{N_{i}}{N_{\gamma}} \, .
  \label{eqn:AlphaDef}
\end{equation}
 The internal conversion process depends on a large number of parameters, including the atomic
 shell involved, $i$, the atomic number, $Z$, the nuclear transition energy, $E_{\gamma}$, and
 on the multipolarity of the transition process, $\tau L$.
 The detailed description of the internal conversion process would go beyond the scope of the present work.
 For further reading we refer to the paper of Band \emph{et al.} \cite{int:2002Ba85} and
 references therein.
 The relativistic expression for  ICC  in the $i-$th atomic sub--shell, derived in the framework of
 the first non-vanishing order of perturbation theory and one-electron approximation for a
 free neutral atom  can be written as:
\begin{equation}
  \alpha^{\tau L}_i\ =\ \sum_{\kappa_f}\left|M^{\tau L}_{i}(\kappa_f)\right|^2 \, ,
  \label{eqn:Alpha}
\end{equation}
 where the partial conversion matrix element  is
 $M^{\tau L}_{i}(\kappa_f)=B^{\tau L}_{i}(\kappa_f)R^{\tau L}_{i}(\kappa_f)$.
 In this expression $B^{\tau L}_{i}(\kappa_f)$ is the angular part,
 $R^{\tau L}_{i}(\kappa_f)$ is the radial part, and $\tau L$ is the nuclear transition multipolarity
 type of electric $\tau=E$ or magnetic $\tau=M$.
 Use is made of relativistic quantum numbers $\kappa =(\ell-j)(2j+1)$, where $\ell$ is
 the electron orbital momentum, and $j$ is the total electron momentum.
 Indices $i$ and $f$ refer to the initial (bound) and final (continuum) states of the electron,
 respectively.
 The summation in Eq.~\ref{eqn:Alpha} extends over all final states allowed by the selection rules
\begin{eqnarray}
  |L-j_i| & \leq  j_f \leq & L+j_i \, ,
\end{eqnarray}
\begin{equation}
    \ell_i+\ell_f+L\,\,\,\rm  =  \left\{     \text{even for $EL$ transitions,}
                                        \atop \text{odd for $ML$ transitions.} \right.
\end{equation}
\\

 For electric transitions,
 $B^{\tau L}_{i}(\kappa_f)$ and $R^{\tau L}_{i}(\kappa_f)$ are given by
 \begin{eqnarray}
  \label{eqn:EL}
   B^{EL}_{i}(\kappa_f) & = & (-1)^{j_f+{\frac12}+L}C^{L0}_{\ell_i0\ell_f 0}W
   \left( \ell_ij_i\ell_{f}j_f;{\scriptstyle\frac12}L\right) \times \nonumber \\
                        &   & \left[\pi k\alpha\frac{(2j_i+1)(2\ell_i+1)
                              (2j_f+1)(2\ell_f+1)}{L(L+1)(2L+1)}\right]^{\frac12}
\end{eqnarray}
 and
 \begin{eqnarray}
  R^{EL}_{i}(\kappa_f) & = & (\kappa_i-\kappa_f)(R_{1,\Lambda=L-1}+R_{2,\Lambda=L-1}) + \nonumber \\
                       &   & L(R_{2,\Lambda=L-1}-R_{1,\Lambda=L-1}+R_{3,\Lambda=L}).
\label{eqn:ELR}
\end{eqnarray}
 In Eq.~\ref{eqn:EL}, $k$ equals $E_{\gamma}$  in units of $m_0c^2$,
 $\alpha$ is the fine structure constant, $C_{\ell_i 0\ell_f 0}^{L0}$ is the Clebsch--Gordan
 coefficient, $W\left(\ell_ij_i\ell_{f}j_f;{\scriptstyle\frac12}L\right)$ is the
 Racah coefficient, and $R_{i,\Lambda}$ is the radial integral (defined below).
 \\

 For magnetic transitions, the corresponding expressions are
 \begin{eqnarray}
 B^{ML}_{i}(\kappa_f) & = & (-1)^{j_f+\frac12+L}C^{L0}_{\ell_i0 \bar\ell_f0}W
   \left(\ell_ij_i\bar\ell_{f}j_f;{\scriptstyle\frac12}L\right) \times \nonumber \\
                      &   & \left[\pi k \alpha\frac{(2j_i+1)(2\ell_i+1)
                      (2j_f+1)(2\bar\ell_f+1)}{L(L+1)(2L+1)}\right]^{\frac12} \, ,
\label{eqn:ML}
\end{eqnarray}
 where $\bar\ell_f=2j_f-\ell_f$, and
\begin{equation}
    R^{ML}_{i}(\kappa_f)\ =\ (\kappa_i+\kappa_f)(R_{1,\Lambda=L}+R_{2,\Lambda=L})\ .
\label{eqn:MLR}
\end{equation}
\\

 Radial integrals in Eqs.~\ref{eqn:ELR} and \ref{eqn:MLR} are written as follows
\begin{equation}
  R_{1,\Lambda}=\int_{0}^{\infty}G_iF_f(E_k)X_\Lambda(kr)dr\,,
  \label{eqn:R1}
\end{equation}
\begin{equation}
  R_{2,\Lambda}=\int_{0}^{\infty}F_iG_f(E_k)X_\Lambda(kr)dr\,,
  \label{eqn:R2}
\end{equation}
\begin{equation}
  R_{3,\Lambda}=\int_{0}^{\infty}[G_iG_f(E_k)+F_iF_f(E_k)] X_\Lambda(kr)dr\,.
  \label{eqn:R3}
\end{equation}
  Functions $G(r)$ and $F(r)$ are given by  $G(r)=rg(r)$ and $F(r)=rf(r)$, where
  $g(r)$ and $f(r)$ are the large and small components of the radial electron wave function,
  respectively.
  In our calculations, $G$ and $F$ are solutions of the Dirac--Fock (DF) equations.
  Wave functions for the bound state $G_i$ and $F_i$ are calculated in the DF field of a neutral atom
  while continuum wave functions $G_f(E_k)$ and $F_f(E_k)$ are determined in the DF field of
  the ion  with a vacancy in the shell from which the conversion electron is emitted.
  In calculating continuum wave functions, the conversion electron energy $E_k$ is determined
  from the energy conservation relationship:
\begin{equation}
  E_k=k-\varepsilon_i\,,
  \label{eqn:Ek}
\end{equation}
 where  $\varepsilon_i$ is the binding energy for the $i-$th atomic sub--shell.
 In the absence of experimental values of $\varepsilon_i$ we use theoretical binding energies for
 ICC calculations of superheavy elements.
 $ X_\Lambda(kr)$ is the transition potential \cite{int:2002Ba85}.
 \\

 The methodology of generating the numerical tables was very similar to our recent work on
 conversion coefficients for the $Z=5-110$ elements \cite{int:2008Ki07}.
 The numerical calculations have been carried out using the RAINE code \cite{int:2002Ba85}.
 Extra care was taken to use a suitable energy mesh for regions where the ICC may change
 dramatically, as in the so--called \emph{``resonance--regions"} \cite{int:2010Tr02}.
 The selection of the input parameters will be discussed further in sections \ref{sec:Mass},
 \ref{sec:BE} and \ref{sec:Vacancies}.

\section{Atomic masses}
\label{sec:Mass}

\begin{figure}[]
\centering
\resizebox{16cm}{!}{\includegraphics{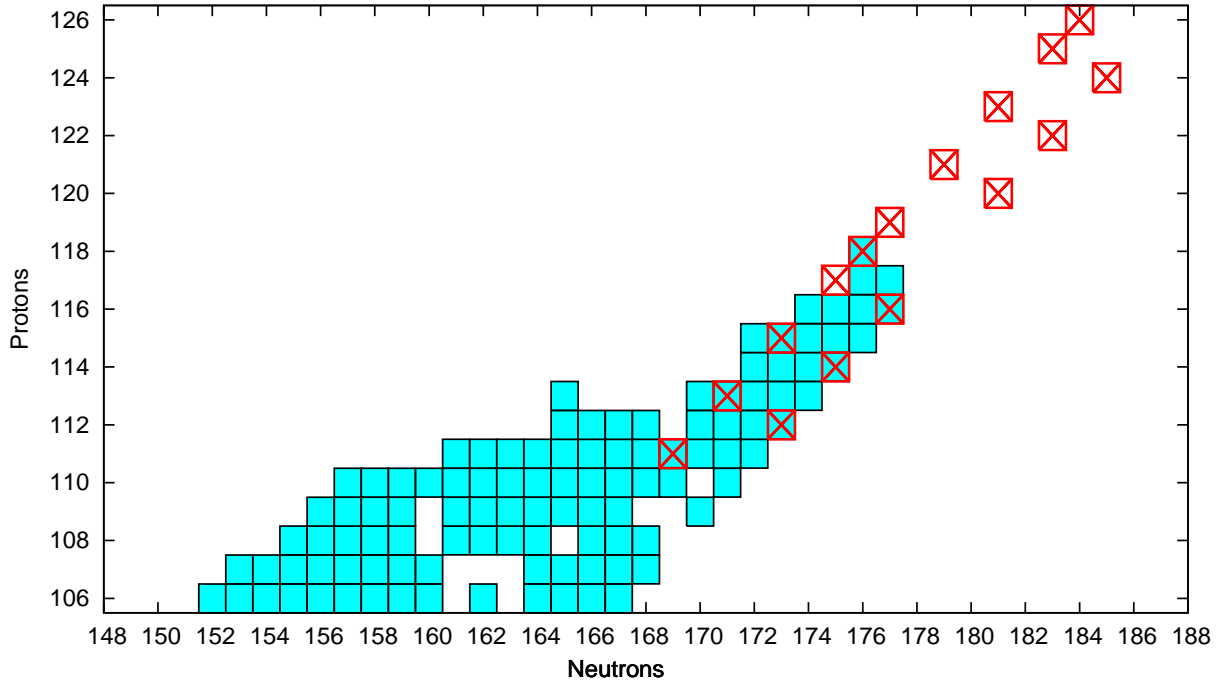}}
\caption{(Color on--line) The chart of nuclides for superheavy elements with $106 \leq Z \leq 118$.
Squares with crosses in red are indicate the isotopes adopted for the present calculations. }
\label{fig:SHE_chart}
\end{figure}

 It is customary to calculate the conversion coefficients for the most abundant isotope
 \cite{int:2002Ba85}.
 For chemical elements with $Z > 92$ no stable isotopes are known, therefore different methods
 must be used to select suitable mass values.
 Fig.~\ref{fig:SHE_chart} shows the partial chart of nuclides with the known isotopes of
 $106 \leq Z \leq 118$ elements.
 The isotopes used for the present ICC calculations are represented by squares marked with red crosses.
 For $Z=111-116$ and $118$ elements we have adopted the recommendations of the Commission on
 Atomic Weights and Isotopic Abundances (IUPAC) \cite{int:AtomicWeights}, which is based
 on the experimental  half--lives and the relative atomic mass values.
 At the time of the last evaluation (2007) no isotope was known for $Z=117$.
 Very recently Oganessian \emph{et al.} \cite{int:2010Og01} have reported the discovery of two new
 $Z=117$ isotopes with $A=293$ and $294$.
 For still heavier elements ($Z>118$) there are no known isotopes.
 Here the varied predictions of ground state properties including the possible locations of shell
 closures at  $Z = 114$, $120$ or $126$  cannot be relied upon to uniquely identify the most likely
 masses above the heaviest known isotope and the elements beyond.
 Thus, for nuclides above $Z=118$, the trend visible in Fig.~\ref{fig:SHE_chart} was extended to the
 higher $Z$ values.
 $Z=126$ is a special case for which a simple linear extrapolation suggests $N=187$ is the corresponding
 isotope.
 However keeping in mind the consistency of theoretical predictions for $N=184$, this was chosen as a good
 candidate for the next $N$-shell closure.
 \\
\begin{table}[]
\centering
\caption{Percentage difference (see Eqn.~\ref{eqn:MassDependence}) of theoretical total ICC`s
 calculated for $N=172$ and $N=184$ neutron numbers and for $Z=126$. }
\label{tab:MassDependence}       %
\begin{tabular}{rccc}
   \hline\noalign{\smallskip}
\multicolumn{1}{c}{$E_{tr}$} &
\multicolumn{1}{c}{Shell} &
\multicolumn{2}{c}{$\Delta \alpha(N172:N184)$ [\%] } \\
\multicolumn{1}{c}{[keV]} &
\multicolumn{1}{c}{ } &
\multicolumn{1}{c}{M1} &
\multicolumn{1}{c}{E2} \\
   \noalign{\smallskip}\hline\noalign{\smallskip}
 100  & K     & $\clubsuit$ & $\clubsuit$ \\
1000  &       & +1.71       & +0.63       \\ \hline
 100  & L     & +1.54       & +0.45       \\
1000  &       & +1.75       & +0.71       \\ \hline
 100  & Total & +1.54       & +0.46       \\
1000  &       & +1.48       & +0.51       \\ \hline
\multicolumn{4}{c}{$\clubsuit$ Energetically not allowed}
\end{tabular}
\end{table}

 The sensitivity of the conversion coefficients to the neutron number has been recently investigated
 in the lighter elements \cite{int:2002Ra45,int:2008Ki07}.
 To explore this sensitivity for heavier systems, calculations for $N=172$ and $N=184$
 are compared for $Z=126$.
 Representative results are provided in Table~\ref{tab:MassDependence} which shows the differences
 between total conversion coefficients calculated for M1 and E2 multipolarities.
 The difference, $\Delta \alpha(N172:N184)$ (in \%) is defined as:
 \begin{equation}
 \label{eqn:MassDependence}
 \Delta \alpha(N172:N184) = \frac{[\alpha(N172) - \alpha(N184)]}{\alpha(N184)} \times 100 \,.
 \end{equation}
 The largest differences of $\sim$1.8\% for M1 transitions is much smaller than the expected 
 accuracy of any realistic conversion coefficient measurement in the region of SHE.

\section{Electron configurations and atomic binding energies}
\label{sec:BE}

 Chemical properties form an important aspect of the characterization of the superheavy elements
 \cite{int:1975Fricke,int:2004Pershina,int:2006Schadel,int:2011Pykko}.
 For the present ICC calculations it was assumed that the emitting nuclide was initially in its
 neutral atomic state.
 In the absence of experimental information on the ground state electron configuration or the
 electron binding energies, the atomic properties were taken from a variety of theoretical
 calculations.
 Table~\ref{tab:AtNucData} lists the adopted ground state configurations for the SHE.
 Most of these calculations incorporate relativistic and electron correlation effects.
 In these heavy systems several configurations could often be very close to each other in terms of
 energy.
 Since these configurations could interact with each other \cite{int:2006Ne11}, multiconfiguration
 Dirac--Fock calculations with the inclusion of a total angular momentum coupling scheme is recommended.
 For example, in the case of $Z=125$ three configurations have been identified \cite{int:2006Ne11}:
 \begin{equation}
    81\% \times [5g_{7/2}^{1} 6f_{5/2}^{2} 8p_{1/2}^{2}] +
    17\% \times [5g_{7/2}^{1} 6f_{5/2}^{1} 7d_{3/2}^{2} 8p_{1/2}^{1}] +
    2\% \times [6f_{5/2}^{2} 6f_{7/2}^{1} 7d_{3/2}^{1} 8p_{1/2}^{1}] \, .
   \label{eqn:Z125conf}
\end{equation}
 Similarly, the ground state of the $Z=126$ element contains two configurations:
 \begin{equation}
    99.8\% \times [5g_{7/2}^{2} 6f_{5/2}^{2} 6f_{7/2}^{1} 8p_{1/2}^{1}] +
    0.2\% \times [5g_{7/2}^{2} 6f_{5/2}^{2} 8p_{1/2}^{2}] \, .
   \label{eqn:Z126conf}
\end{equation}
 The use of multiconfiguration calculations is very important for understanding the valence shells.
 However, for conversion coefficient calculations, which are dominated by contributions from the
 inner shells, it is sufficient to use the most dominant configuration.
\\

 The selection of atomic shell binding energies could have a significant impact on the conversion coefficients,
 especially for cases when the kinetic energy of the emitted electrons is very low.
 It should be also noted that accurate knowledge of the binding energies is required to identify
 conversion electrons and X--ray radiations, and to determine the atomic number of the emitting nucleus.
 \\

\begin{figure}[t]
\centering
\resizebox{16cm}{!}{\includegraphics[angle=-90]{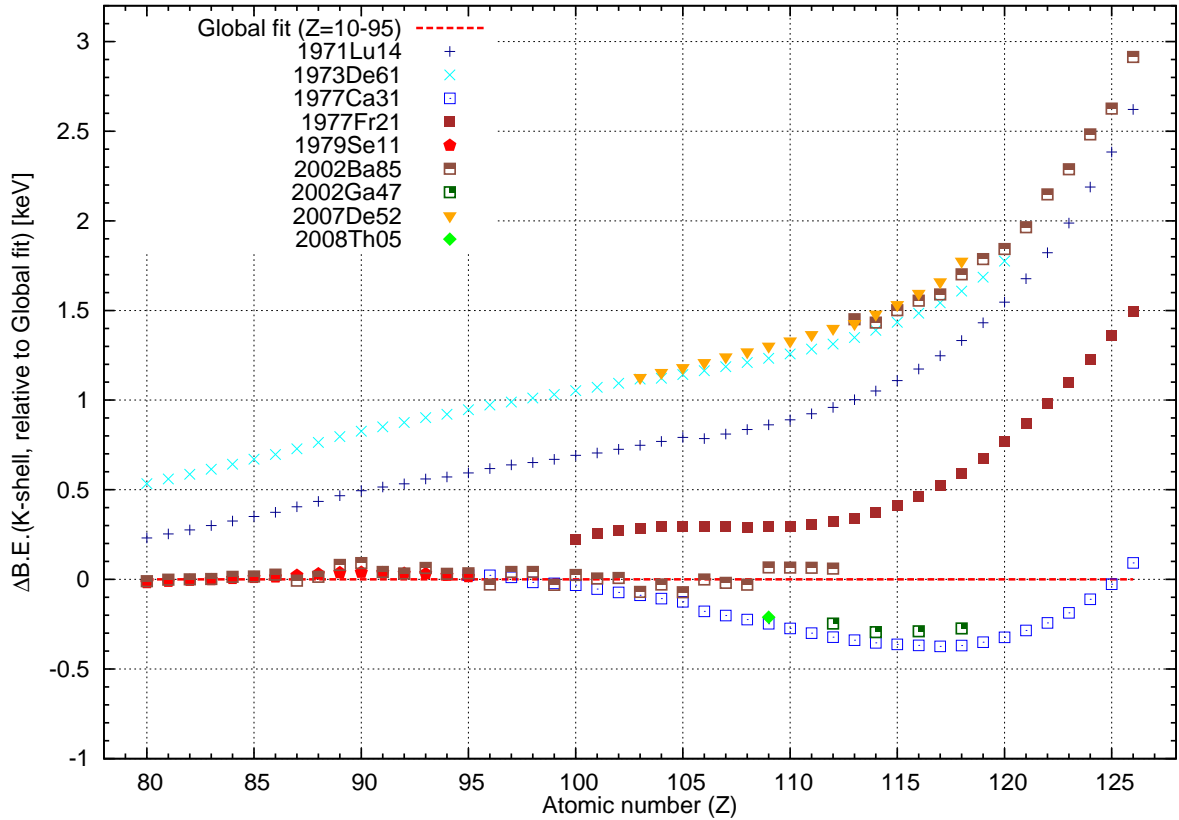}}
\caption{(Color on--line) Relative $K$--shell binding energies for  $80 \leq Z \leq 126$ elements.
Data taken from
1971Lu14 \cite{int:1971Lu14};
1973De61 \cite{int:1973De61};
1977Ca31 \cite{int:1977Ca31};
1977Fr21 \cite{int:1977Fr21};
1979Se11 \cite{int:1979Se11};
2002Ba85 \cite{int:2002Ba85};
2002Ga47 \cite{int:2002Ga47};
2007De52 \cite{int:2007De52} and
2008Th05 \cite{int:2008Th05}.
The red (on-line) dashed line represents a polynomial fit to experimental ($Z=10-95$ \cite{int:1979Se11})
and theoretical ($Z=96-110$ \cite{2002Ba85}) binding energies.
These values were used in the recent ICC calculations \cite{int:2008Ki07}.}
\label{fig:BinEnerDiff}
\end{figure}

 Figure \ref{fig:BinEnerDiff} compares the available data, mainly from theoretical sources, on the
 $K$--shell binding energies of $Z=80$ to $126$ elements.
 It is expected that the $K$--shell binding energies display a regular dependence on the atomic number.
 In fact, the continuous line on the figure is a \emph{``global fit"} to the values adopted for our recent ICC
 calculations for $Z = 5 - 110$ \cite{int:2008Ki07}.
 The values are taken from the compilation of experimental values by Sevier \cite{int:1979Se11}
 (1979Se11, $Z \leq 95$) and
 from the Dirac--Fock theoretical calculations of Band \emph{et al.}, with regard for corrections
 of higher orders, that is the Breit magnetic electron interaction and quantum electrodynamical (QED)
 corrections \cite{int:2002Ba85} (2002Ba85, $Z = 96 - 110$).
 There is excellent agreement between the adopted values and the polynomial fit for $Z = 10 - 110$; the
 largest deviation is less than 0.1 keV.
 Extrapolated values of the \emph{``global fit"} were used to compare the various theoretical calculations above
 $Z=110$.
 A striking feature of the figure is that most calculations overestimate  the $K$--shell binding energies
 compared to the \emph{``global fit"}, even below $Z = 110$ atomic numbers where experimental data exist.
 The earliest calculation of Lu \emph{et al.}, \cite{int:1971Lu14} (1971Lu14) is based on a Dirac---Slater
 model and covers all atomic shells in neutral atoms with $Z=2-126$.
 This table was used in the recent conversion coefficient calculations of
 Ry\v{s}av\'{y} and Dragoun \cite{int:2001Ry04}.
 Desclaux \cite{int:1973De61} (1973De61) carried out relativistic Dirac--Fock calculations for $Z=1-120$.
 A modified version of this model together with empirical corrections has been used by
 Carlson and Nestor \cite{int:1977Ca31} (1973Ca31), and their values are very close to the
 \emph{``global fit"} for $Z \geq 105$ elements.
 Based on Dirac--Slater calculations Fricke and Soff \cite{int:1977Fr21} (1973Fr21) presented detailed
 tables for the $Z=100$ to $173$ superheavy elements.
 While this calculation agrees reasonably well with the \emph{``global fit}" up to about $Z=115$, for heavier systems
 it gives significantly higher values than expected.
 The recent Band \emph{et al.} \cite{int:2002Ba85} ICC calculations for Z=95-126 elements has used
 relativistic Hartree--Fock--Slater (with semi--empirical corrections) ($Z \leq 112$) and relativistic
 Dirac--Fock values ($Z > 112$).
 For $Z>112$ elements their values much higher than the \emph{``global fit"}.
 Gaston, Schwerdtfeger and Nazarewitz \cite{int:2002Ga47} (2002Ga47), using a Dirac--Fock theory taking into account
 the Breit electron interaction and  QED corrections, have calculated the $K$ and $L1$ shell binding energies
 very accurately for $Z=112$, $114$, $116$ and $118$ elements.
 More recently De Macedo \emph{et al.} \cite{int:2007De52} (2007De52) have presented a prolapse free Gaussian basis
 set for relativistic calculations.
 Their results overlap with the results of Carlson and Nestor \cite{int:1977Ca31} and
 Thierfelder \emph{et al.} \cite{int:2008Th05} (2008Th05).
 \\

 Table~\ref{tab:AtNucData} contains the adopted atomic shell binding energies for $K$ ($1s$) to $R2$ ($8p$) shells.
 Based on the level of agreement with the \emph{``global fit"} to the $K$--shell binding energies shown
 in Fig.~\ref{fig:BinEnerDiff}, the theoretical values from Carson and Nestor \cite{int:1977Ca31} have been
 adopted for the present ICC calculations for $Z=111-126$.
 These calculations cover the $K$, $L1-L3$, $M1-M1$, $N1-N5$ and $O2-O4$ shells.
 These shells are usually the most important ones for the conversion process.
 For the remaining shells, values from Fricke and Soff \cite{int:1977Fr21}  were adopted.
 As the binding energies were adopted from different theoretical calculations,
 there might be a small  discontinuity ($\sim 0.3$ keV for the $K$--shell) around $Z=110$.
 Further experimental data as well as improved Dirac--Fock calculations with the Breit and
 QED corrections are needed to improve the atomic data for the ICC calculations and resolve this discrepancy.
 Except for very low transition energies, the above uncertainty in the binding energies has a relatively
 small impact on the total conversion coefficients.

\section{Treatment of the atomic vacancies}
\label{sec:Vacancies}

 \addtolength{\tabcolsep}{-0.10cm}
\begin{table}[]
\centering
\caption{Selected conversion coefficients (ICC) for $Z=126$ obtained from various theoretical calculations.}
\label{tab:Z126ICC}       %
\begin{tabular}{cc|cccc|cccc|}
   \hline\noalign{\smallskip}
\multicolumn{1}{ c }{Shell} &
\multicolumn{1}{ c|}{$E_{tr}$} &
\multicolumn{4}{ c|}{ICC(M1) } &
\multicolumn{4}{ c|}{ICC(E2) }\\
\multicolumn{1}{ c }{} &
\multicolumn{1}{ c|}{[keV]} &
\multicolumn{1}{ c }{NICC } &
\multicolumn{1}{ c }{BrIcc V1.3 } &
\multicolumn{1}{ c }{BrIccNH$^{(a}$} &
\multicolumn{1}{ c|}{BrIccFO$^{(b}$} &
\multicolumn{1}{ c }{NICC } &
\multicolumn{1}{ c }{BrIcc V1.3 } &
\multicolumn{1}{ c }{BrIccNH$^{(a}$ } &
\multicolumn{1}{ c|}{BrIccFO$^{(b}$ } \\
\multicolumn{1}{ c }{} &
\multicolumn{1}{ c|}{} &
\multicolumn{1}{ c}{\cite{int:2001Ry04} } &
\multicolumn{1}{ c}{\cite{int:2002Ba85} } &
\multicolumn{1}{ c }{} &
\multicolumn{1}{ c|}{} &
\multicolumn{1}{ c }{\cite{int:2001Ry04} } &
\multicolumn{1}{ c }{\cite{int:2002Ba85} } &
\multicolumn{1}{ c }{} &
\multicolumn{1}{ c|}{} \\
   \noalign{\smallskip}\hline\noalign{\smallskip}
Total  &  10  & 1.883E+4  & --            & 1.670E+4  & 1.673E+4  & 2.248E+6   & --            & 2.070E+6  & 2.072E+6 \\
   &  50  & 5.128E+2  & --            & 4.647E+2  & 4.650E+2  & 4.555E+3   & --            & 4.368E+3  & 4.374E+3 \\
   & 100  & 2.445E+2  & 2.24E+2       & 2.238E+2  & 2.243E+2  & 3.419E+2   & 3.30E+2       & 3.300E+2  & 3.319E+2 \\
   & 300  & 3.075E+1  & 2.84E+1       & 2.841E+1  & 2.852E+1  & 3.552E+0   & 3.48E+0       & 3.475E+0  & 3.493E+0 \\
   &1000  & 9.807E-1  & 8.82E-1       & 8.814E-1  & 8.817E-1  & 1.236E-1   & 1.19E-1       & 1.193E-1  & 1.199E-1 \\
\end{tabular}
\end{table}
 The atomic vacancy created in the conversion process has a non--negligible effect on the conversion
 process itself.
 In one extreme assumption -- the so--called \emph{``No--Hole"} approximation -- 
 Table \ref{tab:Z126_TotIcc} the vacancy is disregarded.
 In this case the vacancy on the atomic shell is filled immediately and the conversion electron moves
 out of the field of a neutral atom.
 This approximation was used in the  Band \emph{et al.} \cite{int:2002Ba85} and
 Ry\v{s}av\'{y} and Dragoun \cite{int:2001Ry04} ICC tabulations.
 The corresponding ICC calculated here is labeled as \emph{``BrIccNH"}.
 An alternative approach, based on the so--called \emph{``Frozen Orbital"} approximation
 incorporates the presence of the vacancy and is labeled here as \emph{``BrIccFO"}.
 The question of whether or not the atomic vacancy should be included cannot be determined from
 theoretical calculations alone.
 The impact of the vacancies is expected to be significant for low $Z$ and/or highly converted
 transitions and close to the shell binding energy \cite{int:2008KiZV}.
 These differences in the total conversion coefficient can reach up to  10\% for $Z=20$ and E2 and or M4
 multipolarity.
 It was shown \cite{int:2008Ki07} that the effect decreases with atomic number.
 The present calculations confirm this trend.
 Table~\ref{tab:Z126ICC} compares conversion coefficients calculated for $Z=126$  and for M1 and E2 multipolarities.
 It contains ICC values for $n \, s_{1/2}$ shells, starting around 2 keV above the shell binding energy, where
 the effect of the vacancy is expected to be large.
 In general the ICC calculated with the \emph{``Frozen Orbital"} approximation is higher
 than the corresponding value calculated with the \emph{``No--Hole"} approximation.
 For cases from Table \ref{tab:Z126ICC}, the largest difference of $\sim$6\% was obtained for
 $K$--shell and E2 multipolarity at 270 keV transition energy.
 For an M4 transition this difference is $\sim$10.6\%.
 \\

 A number of high precision measurements have been carried out recently to determine the
 conversion coefficient of high multipole order transitions in order to determine the correct treatment of
 the atomic vacancies during the conversion process.
 (See for example \cite{int:2009Ni13} and references therein.)
 Based on our analysis \cite{int:2008KiZV,int:2008Ki07} there is a definite preference towards the
 \emph{``Frozen Orbital"} approximation and this approximation was used to calculate the current ICC tables.
 It should be noted that we are not aware of any suitable experimentally derived conversion coefficient
 which could serve as a basis for deciding which atomic model is best for the superheavy elements.

\section{Comparison with previous ICC calculations}
\label{sec:calc}

\begin{figure}[]
\centering
\resizebox{16cm}{!}{\includegraphics{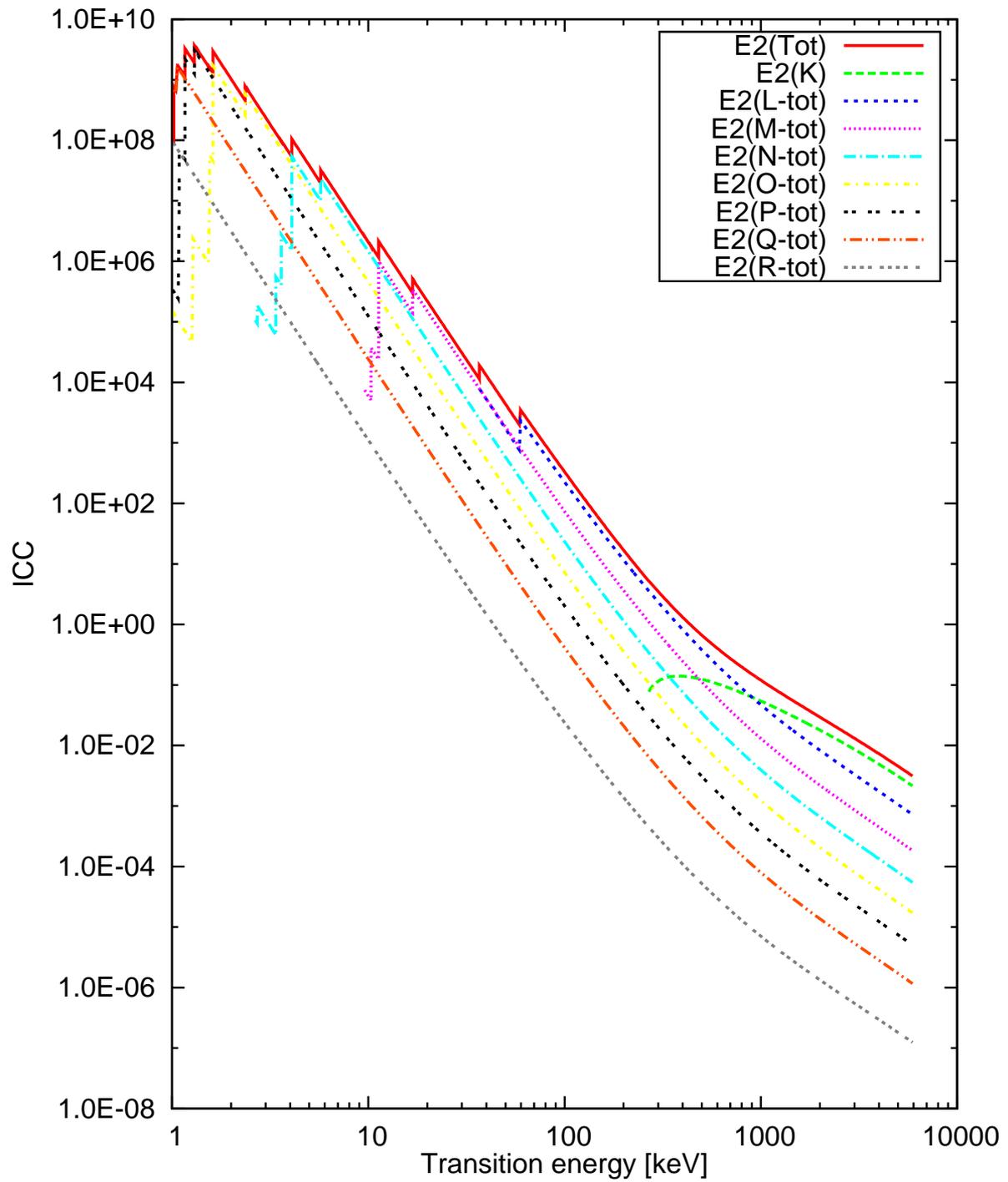}
}
\caption{(Color on--line) Theoretical E2 conversion coefficients calculated for $Z=126$.}
\label{fig:Z126_E2}
\end{figure}
 Representative conversion coefficients for $Z=126$ and pure E2 transitions are shown in
 Fig.~\ref{fig:Z126_E2}.
 The calculations have been carried out for all sub--shells ($K$ to $R2$), but the figure only
 shows the major shells and the total ICC.
 The sharp peaks in the ICCs occur when the transition energy passes one of the boundaries
 located 1 keV above the sub--shell binding energies.
 \\

 There are two existing ICC tabulations for superheavy elements.
 Ry\v{s}av\'{y} and  Dragoun \cite{int:2001Ry04} used the  Dirac--Slater atomic model.
 This table, denoted NICC, contains ICC values for $104 \leq Z \leq 126$ elements for E1--E4 and M1--M4 multipolarities.
 In this table 20 transition energies between 10 keV and 5 MeV were used for the $K$--shell and $L$--subshells.
 For higher shells the calculations only cover the 10 keV to 1000 keV energy region.
 Table~\ref{tab:Z126ICC} compares some representative ICC values of the NICC table \cite{int:2001Ry04} with the
 present calculations.
 The values in the NICC table are systematically larger by 5--10\%.
 \\

 The Band \emph{et al.} \cite{int:2002Ba85} ICC tables (labeled as BrIcc v1.3) use essentially the same
 physical model (relativistic Dirac--Fock) as the present BrIccNH \emph{``No--Hole"} approximation.
 However some of the parameters (binding energies and atomic masses) have been replaced with more
 up--to--date values.
 There is an overall agreement between the Band \emph{et al.} \cite{int:2002Ba85} and the present BrIccNH
 ICC values, however close to the shell binding energies up to $\sim$5\% differences can be seen
 in Table \ref{tab:Z126ICC}.
 The Band \emph{et al.} \cite{int:2002Ba85} calculations cover the E1--E5 and M1--M5 multipolarities, all
 atomic shells, and $10 \leq Z \leq 126$ elements.
 This tabulations starts at 1 keV above the $L1$--shell binding energy and the highest transition energy is 2000 keV.
 \\

 The current tabulations (BrIccFO) use the \emph{``Frozen Orbital"} approximation and the energy range has been
 extended for 1 keV above the atomic binding energies up to 6000 keV.
 Particular attention was paid to use a sufficiently large number of energy mesh points to minimise
 the interpolation error.
 Moreover, to obtain ICC for major shells ($L$, $M$, $N$, etc) and the total conversion coefficient,
 one should obtain the sub--shell values first and then add them together.

\section{Suppression of K-shell conversion in heavy systems}
\label{sec:KLratio}

 A characteristic feature of conversion electron measurements in the heaviest nuclei is the
 absence of the $K$--shell conversion electrons for $L \geq 2$ transitions.
 For example in the recent study on $^{254}$No
 \cite{int:2008HeAA} for the highest energy E2 transition at 267 keV ($10^{+} \rightarrow 8^{+}$)
 the $K/L$ ratio is 0.326.
 For a similar energy transition in $Z=20$ the ratio would be 11.5, which is 35 times larger.
 The $K/L$ ratio, which is often used to determine the transition multipolarity,
 especially at low energy, has a strong dependence on transition energy, multipolarity
 and atomic number.
 At higher energies the $K/L$ ratio is less energy dependent and reaches nearly the same value
 for all multipolarities.
 For selected elements Table \ref{tab:KLratio} shows a list of the $K/L$ ratios for the lowest $L \leq 3$
 multipole orders.
 It is also evident from the table that the $K/L$ ratio values for 6000 keV are in inverse relation to
 the atomic number.\\

\begin{figure}[]
\centering
\resizebox{13cm}{!}{\includegraphics{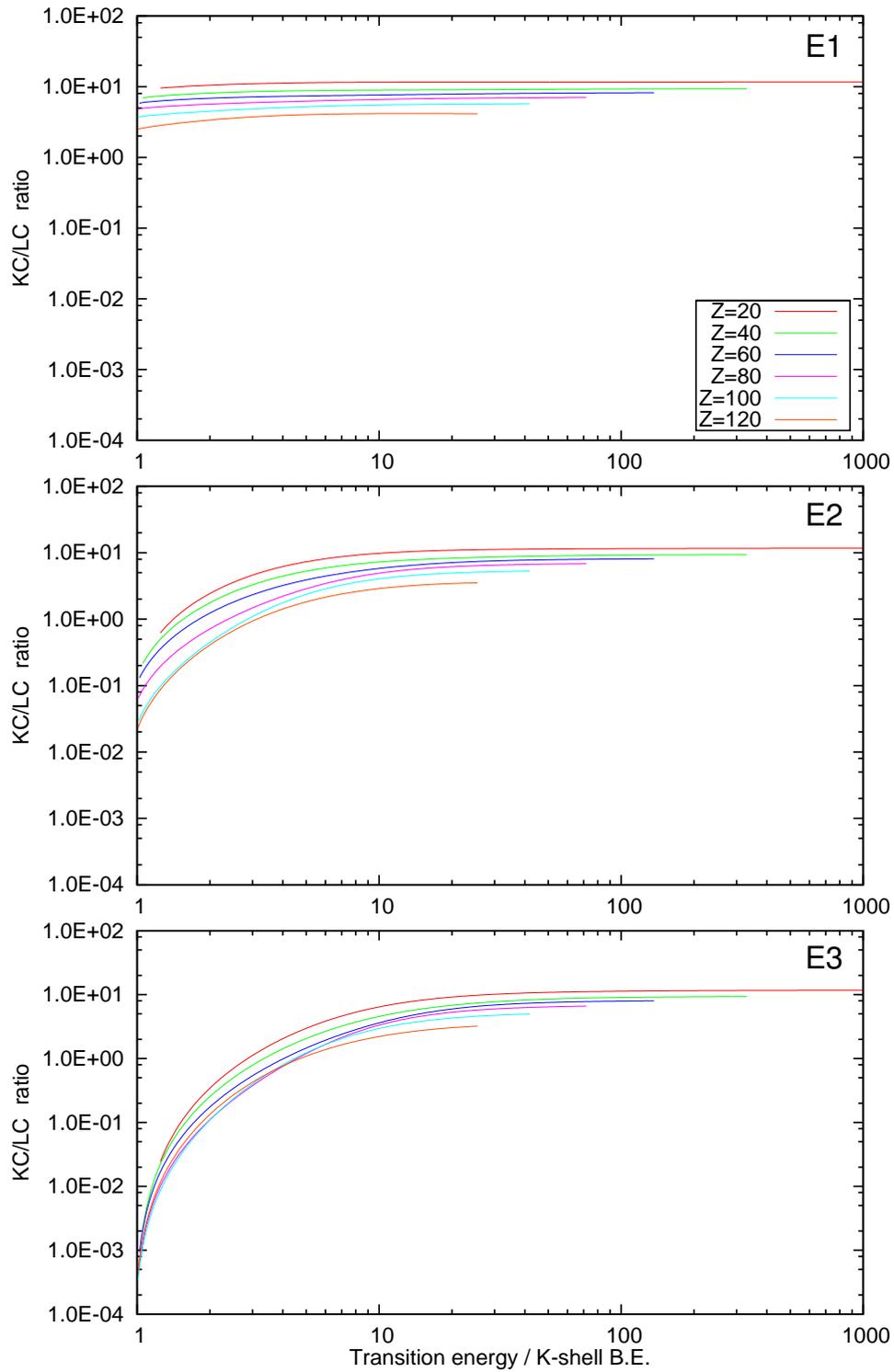}
}
\caption{(Color on--line) Theoretical $K/L$ conversion coefficient ratios for E1, E2 and E3 transitions
 for selected atomic numbers between $Z=20$ and $Z=120$.
 The horizontal axis is the transition energy divided by the $K$--shell binding energies in the
 respective elements.}
\label{fig:KpL_E123}
\end{figure}

\begin{figure}[]
\centering
\resizebox{13cm}{!}{\includegraphics{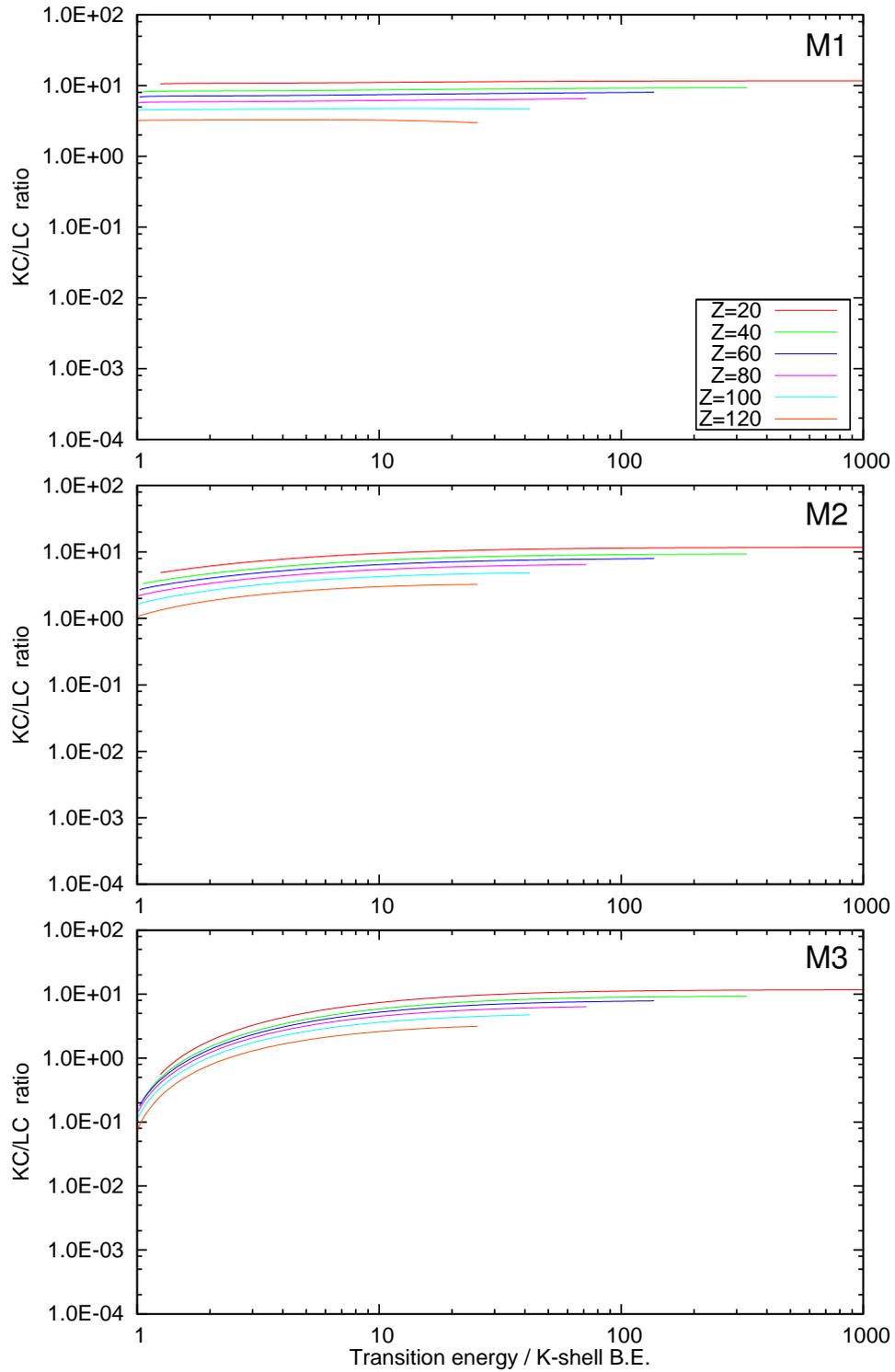}
}
\caption{(Color on--line) Same as Fig.~\ref{fig:KpL_E123} but for M1, M2 and M3 multipolarities.}
\label{fig:KpL_M123}
\end{figure}

\begin{table}[t]
\centering
\caption{Theoretical $K/L$ ICC ratios for selected elements and transition energies.
 For each element the lowest transition energy ($\clubsuit$)
 is about 10 keV above the $K$--shell binding energy.}
\label{tab:KLratio}       %
\begin{tabular}{rr@{}lr@{.}lr@{.}lr@{.}lr@{.}lr@{.}lr@{.}l}
   \hline\noalign{\smallskip}
\multicolumn{1}{c}{Z} &
\multicolumn{2}{c}{$E_{tr}$} &
\multicolumn{12}{c}{K/L ratio} \\
\multicolumn{1}{c}{} &
\multicolumn{2}{c}{[keV]} &
\multicolumn{2}{c}{E1 } &
\multicolumn{2}{c}{E2} &
\multicolumn{2}{c}{E3} &
\multicolumn{2}{c}{M1} &
\multicolumn{2}{c}{M2} &
\multicolumn{2}{c}{M3} \\
   \noalign{\smallskip}\hline\noalign{\smallskip}
 20 &   14 &$\clubsuit$ & 11&15 &  5&41  &  1&55    & 10&88 &  7&46 &  3&78 \\
    & 6000 &            & 11&71 & 11&71  & 11&70    & 11&71 & 11&70 & 11&69 \\ \hline
 50 &   39 &$\clubsuit$ &  6&95 &  0&551 &  0&0333  &  7&72 &  3&56 &  0&590 \\
    & 6000 &            &  8&77 &  8&71  &  8&64    &  8&69 &  8&62 &  8&55  \\ \hline
 80 &   93 &$\clubsuit$ &  5&05 &  0&120 &  0&00340 &  5&85 &  2&39 &  0&264 \\
    & 6000 &            &  7&08 &  6&83  &  6&61    &  6&57 &  6&47 &  6&36  \\ \hline
 110&  193 &$\clubsuit$ &  3&23 &  0&0350&  0&00145 &  3&89 &  1&43 &  0&142 \\
    & 6000 &            &  4&97 &  4&42  &  4&10    &  3&77 &  4&02 &  3&93  \\ \hline
 126&  278 &$\clubsuit$ &  2&23 &  0&0293&  0&00104 &  2&84 &  0&951&  0&0855 \\
    & 6000 &            &  3&62 &  2&99  &  2&70    &  2&57 &  2&80 &  2&70  \\ \hline
\end{tabular}
\end{table}

 Considering the dependence of the $K/L$ ratio on the various parameters Listengarten
 \cite{int:1961Listengarten} has emphasized that this ratio for E1 and M1 transitions
 is defined mainly by the atomic number $Z$ and its value decreases from $9-10$ at $Z=33$
  to $4-5$ at $Z=98$.
  For the higher multipoles, the $K/L$ value depends in the first  approximation only on the
  $E_{tr}/\varepsilon_K$ ratio,  that is, on $E_{tr}/Z^{2}$ (because $\varepsilon_{K} \propto Z^{2}$
  in the Coulomb approximation, where $\varepsilon_{K}$ is the $K$--shell binding energy).
  Therefore, the ratio decreases as $Z$ increases.
  For high $Z$ and for high multipolarities the $K/L$ ratio is much smaller than for E1 and M1 and may be
  even less than 1 in the practically important energy ranges.
  At close to the threshold energy, the $K/L$ ratio for E5 transitions may decrease to $10^{-5}$."
  This is illustrated in Figs.~\ref{fig:KpL_E123} and in \ref{fig:KpL_M123}, which show the $K/L$ ratios for the
  lowest 3 electric and magnetic multipole orders.
  By replacing the transition energy with $E_{tr}/\varepsilon_{K}$ all curves are largely shifted on
  top of each other, i.e. the $K/L$ ratios for a given multipolarity
  and for different atomic numbers have similar shape, even for the highest $Z$ values.

\section{Data tables and graphs}
\label{sec:Data tables}

 The ICC`s were calculated for elements of $111 \leq Z \leq 126$.
 The relevant atomic data, including atomic mass, electron configurations and
 atomic binding energies, which are compiled from a number of sources, are given in
 Table~\ref{tab:AtNucData}.
 \\

 The ICC table calculated here for all atomic shells is around 47000 lines long.
 Here we present only tables of the total conversion coefficients.
 Tables~\ref{tab:Z111_TotIcc} to \ref{tab:Z126_TotIcc} show the values for E1--E5 and M1--M5
 multipolarities.
 These values have been obtained by numerical interpolation of the sub--shell values.
 As the tabulations start 1 keV above the binding energies, the total conversion coefficients
 only contain contributions above these thresholds, which are indicted by the relevant atomic shells
 in the second column.
 The complete table for ICC values for all sub--shells is available from the website at
 www.academicpress.com/adndt.
 The new ICC data table for superheavy elements calculated here will be incorporated
 in the next release of the BrIcc \cite{int:2008Ki07} data base.

%% file: TableExplanation.tex
\clearpage
\newpage

\TableExplanation

\bigskip
\renewcommand{\arraystretch}{1.0}
\section*{Table 1. \label{sec:AtNucData} Adopted atomic and nuclear data used for
          the calculations.}
\begin{tabular*}{0.95\textwidth}{@{}@{\extracolsep{\fill}}lp{5.5in}@{}}
\multicolumn{2}{p{0.95\textwidth}}{For each element the adopted atomic weight and electron
    configuration above the configuration of Rn ($Z=86$) is given.
    Atomic weight for $111 \leq Z \leq 118$ are taken from \cite{AtomicWeights} and a
    linear extrapolation was used for heavier elements.
    The electron configurations for $111 \leq Z \leq 118$ are taken from \cite{2002Ba85} and
    from \cite{2006Ne11} for heavier elements.}\\

Shell	&
Major atomic shell\\

Binding energy &
    Binding energies for the
    $K$, $L1-L3$, $M1-M5$, $N1-N5$ and $O2-O4$ shells from the relativistic  Dirac--Fock calculations
    of Carlson and Nestor \cite{1977Ca31}
    and for the other shells from the Dirac--Slater calculations of
    Fricke and Soff \cite{1977Fr21}.
    Energies are given in keV. \\
Reference & Citation to the electron configurations. \\
\end{tabular*}
\section*{Table 2.  Total conversion coefficients for Z=111}
\begin{tabular*}{0.95\textwidth}{@{}@{\extracolsep{\fill}}lp{5.5in}@{}}
\multicolumn{2}{p{0.95\textwidth}}{Total conversion coefficients for all elements
 evaluated from partial values using BrIcc \cite{2008Ki07}.  \label{sec:TotIcc} }\\

Transition energy  & Transition energies given in keV.\\
Shell            & Indicating the energy of the starting point of the sub--shell contributions to the total ICC. \\
ICC              & Total conversion coefficients for the lowest 5 electric (E1--E5) and magnetic (M1--M5)
                   multipole orders\\
\end{tabular*}
\section*{Table 3.  Total conversion coefficients for $Z=112$}
\begin{tabular*}{0.95\textwidth}{@{}@{\extracolsep{\fill}}lp{5.5in}@{}}
\multicolumn{2}{p{0.95\textwidth}}{Same as for Table 2.}\\
\end{tabular*}
\section*{Table 4.  Total conversion coefficients for $Z=113$}
\begin{tabular*}{0.95\textwidth}{@{}@{\extracolsep{\fill}}lp{5.5in}@{}}
\multicolumn{2}{p{0.95\textwidth}}{Same as for Table 2.}\\
\end{tabular*}
\section*{Table 5.  Total conversion coefficients for $Z=114$}
\begin{tabular*}{0.95\textwidth}{@{}@{\extracolsep{\fill}}lp{5.5in}@{}}
\multicolumn{2}{p{0.95\textwidth}}{Same as for Table 2.}\\
\end{tabular*}
\section*{Table 6.  Total conversion coefficients for $Z=115$}
\begin{tabular*}{0.95\textwidth}{@{}@{\extracolsep{\fill}}lp{5.5in}@{}}
\multicolumn{2}{p{0.95\textwidth}}{Same as for Table 2.}\\
\end{tabular*}
\section*{Table 7.  Total conversion coefficients for $Z=116$}
\begin{tabular*}{0.95\textwidth}{@{}@{\extracolsep{\fill}}lp{5.5in}@{}}
\multicolumn{2}{p{0.95\textwidth}}{Same as for Table 2.}\\
\end{tabular*}
\section*{Table 8.  Total conversion coefficients for $Z=117$}
\begin{tabular*}{0.95\textwidth}{@{}@{\extracolsep{\fill}}lp{5.5in}@{}}
\multicolumn{2}{p{0.95\textwidth}}{Same as for Table 2.}\\
\end{tabular*}
\section*{Table 9.  Total conversion coefficients for $Z=118$}
\begin{tabular*}{0.95\textwidth}{@{}@{\extracolsep{\fill}}lp{5.5in}@{}}
\multicolumn{2}{p{0.95\textwidth}}{Same as for Table 2.}\\
\end{tabular*}
\section*{Table 10.  Total conversion coefficients for $Z=119$}
\begin{tabular*}{0.95\textwidth}{@{}@{\extracolsep{\fill}}lp{5.5in}@{}}
\multicolumn{2}{p{0.95\textwidth}}{Same as for Table 2.}\\
\end{tabular*}
\section*{Table 11.  Total conversion coefficients for $Z=120$}
\begin{tabular*}{0.95\textwidth}{@{}@{\extracolsep{\fill}}lp{5.5in}@{}}
\multicolumn{2}{p{0.95\textwidth}}{Same as for Table 2.}\\
\end{tabular*}
\section*{Table 12.  Total conversion coefficients for $Z=121$}
\begin{tabular*}{0.95\textwidth}{@{}@{\extracolsep{\fill}}lp{5.5in}@{}}
\multicolumn{2}{p{0.95\textwidth}}{Same as for Table 2.}\\
\end{tabular*}
\section*{Table 13.  Total conversion coefficients for $Z=122$}
\begin{tabular*}{0.95\textwidth}{@{}@{\extracolsep{\fill}}lp{5.5in}@{}}
\multicolumn{2}{p{0.95\textwidth}}{Same as for Table 2.}\\
\end{tabular*}
\section*{Table 14.  Total conversion coefficients for $Z=123$}
\begin{tabular*}{0.95\textwidth}{@{}@{\extracolsep{\fill}}lp{5.5in}@{}}
\multicolumn{2}{p{0.95\textwidth}}{Same as for Table 2.}\\
\end{tabular*}
\section*{Table 15.  Total conversion coefficients for $Z=124$}
\begin{tabular*}{0.95\textwidth}{@{}@{\extracolsep{\fill}}lp{5.5in}@{}}
\multicolumn{2}{p{0.95\textwidth}}{Same as for Table 2.}\\
\end{tabular*}
\section*{Table 16.  Total conversion coefficients for $Z=125$}
\begin{tabular*}{0.95\textwidth}{@{}@{\extracolsep{\fill}}lp{5.5in}@{}}
\multicolumn{2}{p{0.95\textwidth}}{Same as for Table 2.}\\
\end{tabular*}
\section*{Table 17.  Total conversion coefficients for $Z=126$}
\begin{tabular*}{0.95\textwidth}{@{}@{\extracolsep{\fill}}lp{5.5in}@{}}
\multicolumn{2}{p{0.95\textwidth}}{Same as for Table 2.}\\
\end{tabular*}

%% file: AtomicNuclearData.tex
\renewcommand{\arraystretch}{1.0}


\setlength{\LTleft}{0pt}
\setlength{\LTright}{0pt}


\setlength{\tabcolsep}{0.5\tabcolsep}

\renewcommand{\arraystretch}{1.0}

\footnotesize 

\begin{longtable}{@{\extracolsep\fill}cllllllllll}
\caption{Adopted atomic and nuclear data used for the calculations.
  See page \pageref{sec:AtNucData} for Explanation of Tables.
  Energies are given in keV.}
  \label{tab:AtNucData}
Shell & 1 & 2 & 3 & 4 & 5 & 6 & 7 & 8 & Reference\\
\hline \\
\endfirsthead
\caption[]{Continued.}
Shell & 1 & 2 & 3 & 4 & 5 & 6 & 7 & 8 & Reference\\
\hline \\
\endhead
\multicolumn{10}{l}{ } \\
\multicolumn{2}{l}{ 111--Rg--280} &
\multicolumn{7}{l}{  $ 5f_{5/2}^{ 6}$ $ 5f_{7/2}^{ 8}$ $ 6d_{3/2}^{ 4}$ $ 6d_{5/2}^{ 6}$ $ 7s_{1/2}^{ 1}$} &
\multicolumn{1}{l}{\cite{2002Ba85}}\\
 K   & 186.136       &               &               &               &               &               &               &               \\ 
 L   &\phe37.881     &\phe36.799     &\phe26.648     &               &               &               &               &               \\ 
 M   &\phe10.187     & \phk9.687     & \phk7.224     & \phk6.447     & \phk6.025     &               &               &               \\ 
 N   & \phk2.926     & \phk2.673     & \phk1.978     & \phk1.597     & \phk1.483     & \phk0.9604    & \phk0.9267    &               \\ 
 O   & \phk0.7283    & \phk0.645     & \phk0.281     & \phk0.251     & \phk0.2445    & \phk0.0671    & \phk0.0614    &               \\ 
 P   & \phk0.1267    & \phk0.0901    & \phk0.0498    & \phk0.0094    & \phk0.0066    &               &               &               \\
\end{longtable}

%% file: Z111_TotIcc.TEX
\renewcommand{\arraystretch}{1.0}


\setlength{\LTleft}{0pt}
\setlength{\LTright}{0pt}


\setlength{\tabcolsep}{0.5\tabcolsep}

\renewcommand{\arraystretch}{1.0}

\footnotesize 

\begin{longtable}{@{\extracolsep\fill}rlcccccccccc}
\caption[Total conversion coefficients for $Z=111$]{Total conversion coefficients for $Z=111$ (Rg--$280$, Roentgenium).
  See page \pageref{sec:TotIcc} for Explanation of Tables.
  \label{tab:Z111_TotIcc}}
E$_{\gamma}$ & Shell & E1 & M1 & E2 & M2 & E3 & M3 & E4 & M4 & E5 & M5 \\
\hline \\
\endfirsthead
\caption[]{Continued.}
E$_{\gamma}$ & Shell & E1 & M1 & E2 & M2 & E3 & M3 & E4 & M4 & E5 & M5 \\
\hline \\
\endhead
   1.0066 & P5 & 6.109E-01 & 2.441E+06 & 9.220E+12 & 1.187E+18 & 1.573E+22
               & 1.016E+01 & 7.126E+05 & 1.613E+13 & 1.081E+19 & 7.210E+23 \\
   1.0094 & Q1 & 1.338E+00 & 1.231E+07 & 1.359E+13 & 2.247E+18 & 3.315E+22
               & 1.120E+04 & 6.900E+08 & 2.965E+13 & 1.063E+19 & 6.995E+23 \\
   1.0497 &    & 1.372E+00 & 1.013E+07 & 1.034E+13 & 1.584E+18 & 2.177E+22
               & 9.968E+03 & 5.680E+08 & 2.258E+13 & 7.480E+18 & 4.562E+23 \\
   1.0498 & P3 & 1.473E+00 & 5.693E+08 & 2.263E+13 & 1.629E+18 & 2.177E+22
               & 1.013E+04 & 2.928E+09 & 6.332E+14 & 2.830E+19 & 5.981E+23 \\
   1.0613 &    & 1.477E+00 & 5.392E+08 & 2.098E+13 & 1.478E+18 & 1.937E+22
               & 9.810E+03 & 2.774E+09 & 5.869E+14 & 2.567E+19 & 5.313E+23 \\
   1.0614 & O7 & 6.226E+01 & 5.400E+08 & 2.102E+13 & 1.644E+18 & 6.928E+22
               & 9.820E+03 & 2.772E+09 & 5.865E+14 & 2.585E+19 & 7.467E+23 \\
   1.0670 &    & 6.180E+01 & 5.260E+08 & 2.026E+13 & 1.569E+18 & 6.542E+22
               & 9.667E+03 & 2.701E+09 & 5.654E+14 & 2.466E+19 & 7.051E+23 \\
   1.0671 & O6 & 1.067E+02 & 5.267E+08 & 2.040E+13 & 1.604E+18 & 9.591E+22
               & 9.683E+03 & 2.700E+09 & 5.655E+14 & 2.475E+19 & 7.067E+23 \\
\end{longtable}

%% file: Z112_TotIcc.tex
\renewcommand{\arraystretch}{1.0}


\setlength{\LTleft}{0pt}
\setlength{\LTright}{0pt}


\setlength{\tabcolsep}{0.5\tabcolsep}

\renewcommand{\arraystretch}{1.0}

\footnotesize 

\begin{longtable}{@{\extracolsep\fill}rlcccccccccc}
\caption[Total conversion coefficients for $Z=112$]{Total conversion coefficients for $Z=112$ (Cp--$285$, Copernicium).
  See page \pageref{sec:TotIcc} for Explanation of Tables.
  \label{tab:Z112_TotIcc}}
E$_{\gamma}$ & Shell & E1 & M1 & E2 & M2 & E3 & M3 & E4 & M4 & E5 & M5 \\
\hline \\
\endfirsthead
\caption[]{Continued.}
E$_{\gamma}$ & Shell & E1 & M1 & E2 & M2 & E3 & M3 & E4 & M4 & E5 & M5 \\
\hline \\
\endhead
   1.0073 & P5 & 6.480E-01 & 2.852E+06 & 1.133E+13 & 1.458E+18 & 1.961E+22
               & 1.185E+01 & 8.505E+05 & 2.015E+13 & 1.362E+19 & 9.201E+23 \\
   1.0098 & Q1 & 1.253E+00 & 5.392E+06 & 1.116E+13 & 1.426E+18 & 1.909E+22
               & 1.373E+04 & 7.912E+08 & 3.211E+13 & 1.337E+19 & 8.955E+23 \\
   1.0103 &    & 1.253E+00 & 5.378E+06 & 1.112E+13 & 1.419E+18 & 1.899E+22
               & 1.371E+04 & 7.893E+08 & 3.200E+13 & 1.331E+19 & 8.907E+23 \\
   1.0104 & P4 & 1.463E+00 & 1.485E+07 & 1.649E+13 & 2.736E+18 & 4.104E+22
               & 1.374E+04 & 8.440E+08 & 3.675E+13 & 1.335E+19 & 8.898E+23 \\
   1.0527 &    & 1.498E+00 & 1.211E+07 & 1.238E+13 & 1.897E+18 & 2.640E+22
               & 1.215E+04 & 6.883E+08 & 2.763E+13 & 9.243E+18 & 5.685E+23 \\
   1.0528 & P3 & 1.616E+00 & 6.364E+08 & 2.591E+13 & 1.947E+18 & 2.641E+22
               & 1.233E+04 & 3.369E+09 & 7.222E+14 & 3.311E+19 & 7.335E+23 \\
   1.0689 &    & 1.619E+00 & 5.900E+08 & 2.331E+13 & 1.700E+18 & 2.244E+22
               & 1.179E+04 & 3.124E+09 & 6.497E+14 & 2.891E+19 & 6.219E+23 \\
   1.0690 & O7 & 6.334E+01 & 5.908E+08 & 2.335E+13 & 1.883E+18 & 7.671E+22
               & 1.180E+04 & 3.122E+09 & 6.493E+14 & 2.910E+19 & 8.616E+23 \\
\end{longtable}

%% file: Z113_TotIcc.tex
\renewcommand{\arraystretch}{1.0}


\setlength{\LTleft}{0pt}
\setlength{\LTright}{0pt}


\setlength{\tabcolsep}{0.5\tabcolsep}

\renewcommand{\arraystretch}{1.0}

\footnotesize 

\begin{longtable}{@{\extracolsep\fill}rlcccccccccc}
\caption[Total conversion coefficients for $Z=113$]{Total conversion coefficients for $Z=113$ (Ut--$284$, Element $113$).
  See page \pageref{sec:TotIcc} for Explanation of Tables.
  \label{tab:Z113_TotIcc}}
E$_{\gamma}$ & Shell & E1 & M1 & E2 & M2 & E3 & M3 & E4 & M4 & E5 & M5 \\
\hline \\
\endfirsthead
\caption[]{Continued.}
E$_{\gamma}$ & Shell & E1 & M1 & E2 & M2 & E3 & M3 & E4 & M4 & E5 & M5 \\
\hline \\
\endhead
   1.0049 & Q2 & 3.916E-01 & 9.289E+07 & 3.588E+12 & 2.254E+16 & 2.219E+19
               & 1.932E+03 & 1.862E+07 & 7.017E+10 & 5.228E+13 & 8.863E+15 \\
   1.0101 & P5 & 1.076E+00 & 9.386E+07 & 1.731E+13 & 1.793E+18 & 2.411E+22
               & 1.916E+03 & 1.916E+07 & 2.512E+13 & 1.697E+19 & 1.157E+24 \\
   1.0115 &    & 1.076E+00 & 9.321E+07 & 1.714E+13 & 1.771E+18 & 2.375E+22
               & 1.908E+03 & 1.903E+07 & 2.487E+13 & 1.676E+19 & 1.139E+24 \\
   1.0116 & Q1 & 1.793E+00 & 9.644E+07 & 1.716E+13 & 1.769E+18 & 2.372E+22
               & 1.912E+04 & 1.004E+09 & 4.028E+13 & 1.681E+19 & 1.138E+24 \\
   1.0137 &    & 1.787E+00 & 9.546E+07 & 1.691E+13 & 1.737E+18 & 2.320E+22
               & 1.900E+04 & 9.932E+08 & 3.970E+13 & 1.650E+19 & 1.113E+24 \\
   1.0138 & P4 & 1.994E+00 & 1.066E+08 & 2.309E+13 & 3.278E+18 & 4.939E+22
               & 1.903E+04 & 1.059E+09 & 4.537E+13 & 1.655E+19 & 1.112E+24 \\
   1.0585 &    & 1.960E+00 & 8.602E+07 & 1.709E+13 & 2.229E+18 & 3.105E+22
               & 1.673E+04 & 8.550E+08 & 3.360E+13 & 1.124E+19 & 6.940E+23 \\
   1.0586 & P3 & 2.097E+00 & 7.780E+08 & 3.177E+13 & 2.284E+18 & 3.106E+22
               & 1.692E+04 & 3.882E+09 & 8.131E+14 & 3.808E+19 & 8.812E+23 \\
\end{longtable}

%% file: Z114_TotIcc.tex
\renewcommand{\arraystretch}{1.0}


\setlength{\LTleft}{0pt}
\setlength{\LTright}{0pt}


\setlength{\tabcolsep}{0.5\tabcolsep}

\renewcommand{\arraystretch}{1.0}

\footnotesize 

\begin{longtable}{@{\extracolsep\fill}rlcccccccccc}
\caption[Total conversion coefficients for $Z=114$]{Total conversion coefficients for $Z=114$ (Uq--$289$, Element $114$).
  See page \pageref{sec:TotIcc} for Explanation of Tables.
  \label{tab:Z114_TotIcc}}
E$_{\gamma}$ & Shell & E1 & M1 & E2 & M2 & E3 & M3 & E4 & M4 & E5 & M5 \\
\hline \\
\endfirsthead
\caption[]{Continued.}
E$_{\gamma}$ & Shell & E1 & M1 & E2 & M2 & E3 & M3 & E4 & M4 & E5 & M5 \\
\hline \\
\endhead
   1.0057 & Q2 & 5.025E-01 & 1.216E+08 & 4.731E+12 & 3.019E+16 & 3.048E+19
               & 2.644E+03 & 2.497E+07 & 9.498E+10 & 7.231E+13 & 1.234E+16 \\
   1.0130 & P5 & 1.214E+00 & 1.211E+08 & 2.110E+13 & 2.138E+18 & 2.901E+22
               & 2.604E+03 & 2.527E+07 & 3.056E+13 & 2.072E+19 & 1.424E+24 \\
   1.0134 &    & 1.214E+00 & 1.208E+08 & 2.104E+13 & 2.131E+18 & 2.888E+22
               & 2.601E+03 & 2.522E+07 & 3.048E+13 & 2.065E+19 & 1.418E+24 \\
   1.0135 & Q1 & 2.053E+00 & 1.249E+08 & 2.105E+13 & 2.129E+18 & 2.885E+22
               & 2.378E+04 & 1.229E+09 & 4.944E+13 & 2.071E+19 & 1.417E+24 \\
   1.0172 &    & 2.041E+00 & 1.226E+08 & 2.053E+13 & 2.061E+18 & 2.774E+22
               & 2.352E+04 & 1.207E+09 & 4.820E+13 & 2.005E+19 & 1.361E+24 \\
   1.0173 & P4 & 2.241E+00 & 1.356E+08 & 2.756E+13 & 3.842E+18 & 5.846E+22
               & 2.355E+04 & 1.286E+09 & 5.489E+13 & 2.011E+19 & 1.360E+24 \\
   1.0645 &    & 2.182E+00 & 1.083E+08 & 2.009E+13 & 2.562E+18 & 3.589E+22
               & 2.057E+04 & 1.027E+09 & 4.003E+13 & 1.339E+19 & 8.288E+23 \\
   1.0646 & P3 & 2.340E+00 & 8.737E+08 & 3.597E+13 & 2.621E+18 & 3.589E+22
               & 2.077E+04 & 4.429E+09 & 9.120E+14 & 4.346E+19 & 1.040E+24 \\
\end{longtable}

%% file: Z115_TotIcc.tex
\renewcommand{\arraystretch}{1.0}


\setlength{\LTleft}{0pt}
\setlength{\LTright}{0pt}


\setlength{\tabcolsep}{0.5\tabcolsep}

\renewcommand{\arraystretch}{1.0}

\footnotesize 

\begin{longtable}{@{\extracolsep\fill}rlcccccccccc}
\caption[Total conversion coefficients for $Z=115$]{Total conversion coefficients for $Z=115$ (Up--$288$, Element $115$).
  See page \pageref{sec:TotIcc} for Explanation of Tables.
  \label{tab:Z115_TotIcc}}
E$_{\gamma}$ & Shell & E1 & M1 & E2 & M2 & E3 & M3 & E4 & M4 & E5 & M5 \\
\hline \\
\endfirsthead
\caption[]{Continued.}
E$_{\gamma}$ & Shell & E1 & M1 & E2 & M2 & E3 & M3 & E4 & M4 & E5 & M5 \\
\hline \\
\endhead
   1.0038 & Q3 & 1.995E-02 & 7.309E+07 & 1.681E+12 & 7.203E+15 & 5.115E+18
               & 1.737E+01 & 3.303E+08 & 9.556E+13 & 3.748E+18 & 3.022E+22 \\
   1.0078 & Q2 & 6.852E-01 & 2.366E+08 & 8.084E+12 & 4.868E+16 & 4.804E+19
               & 3.795E+03 & 3.585E+08 & 9.308E+13 & 3.617E+18 & 2.896E+22 \\
   1.0168 &    & 6.585E-01 & 2.264E+08 & 7.601E+12 & 4.503E+16 & 4.393E+19
               & 3.696E+03 & 3.430E+08 & 8.749E+13 & 3.341E+18 & 2.632E+22 \\
   1.0169 & Q1 & 1.643E+00 & 2.314E+08 & 7.632E+12 & 4.503E+16 & 4.390E+19
               & 2.998E+04 & 1.819E+09 & 1.108E+14 & 3.438E+18 & 2.637E+22 \\
   1.0176 &    & 1.638E+00 & 2.306E+08 & 7.596E+12 & 4.476E+16 & 4.359E+19
               & 2.992E+04 & 1.813E+09 & 1.102E+14 & 3.417E+18 & 2.618E+22 \\
   1.0177 & P5 & 2.398E+00 & 2.347E+08 & 2.711E+13 & 2.499E+18 & 3.395E+22
               & 2.993E+04 & 1.813E+09 & 1.465E+14 & 2.814E+19 & 1.732E+24 \\
   1.0226 &    & 2.378E+00 & 2.292E+08 & 2.622E+13 & 2.394E+18 & 3.224E+22
               & 2.950E+04 & 1.770E+09 & 1.417E+14 & 2.695E+19 & 1.643E+24 \\
   1.0227 & P4 & 2.576E+00 & 2.445E+08 & 3.423E+13 & 4.446E+18 & 6.798E+22
               & 2.954E+04 & 1.865E+09 & 1.496E+14 & 2.702E+19 & 1.642E+24 \\
\end{longtable}

%% file: Z116_TotIcc.tex
\renewcommand{\arraystretch}{1.0}


\setlength{\LTleft}{0pt}
\setlength{\LTright}{0pt}


\setlength{\tabcolsep}{0.5\tabcolsep}

\renewcommand{\arraystretch}{1.0}

\footnotesize 

\begin{longtable}{@{\extracolsep\fill}rlcccccccccc}
\caption[Total conversion coefficients for $Z=116$]{Total conversion coefficients for $Z=116$ (Uh--$293$, Element $116$).
  See page \pageref{sec:TotIcc} for Explanation of Tables.
  \label{tab:Z116_TotIcc}}
E$_{\gamma}$ & Shell & E1 & M1 & E2 & M2 & E3 & M3 & E4 & M4 & E5 & M5 \\
\hline \\
\endfirsthead
\caption[]{Continued.}
E$_{\gamma}$ & Shell & E1 & M1 & E2 & M2 & E3 & M3 & E4 & M4 & E5 & M5 \\
\hline \\
\endhead
   1.0044 & Q3 & 2.970E-02 & 1.058E+08 & 2.398E+12 & 1.034E+16 & 7.506E+18
               & 2.417E+01 & 4.848E+08 & 1.410E+14 & 5.589E+18 & 4.590E+22 \\
   1.0101 & Q2 & 8.635E-01 & 3.148E+08 & 1.062E+13 & 6.439E+16 & 6.477E+19
               & 5.133E+03 & 5.170E+08 & 1.357E+14 & 5.314E+18 & 4.319E+22 \\
   1.0202 &    & 8.264E-01 & 2.996E+08 & 9.916E+12 & 5.901E+16 & 5.859E+19
               & 4.982E+03 & 4.920E+08 & 1.266E+14 & 4.863E+18 & 3.881E+22 \\
   1.0203 & Q1 & 1.955E+00 & 3.057E+08 & 9.954E+12 & 5.901E+16 & 5.854E+19
               & 3.681E+04 & 2.260E+09 & 1.545E+14 & 4.980E+18 & 3.887E+22 \\
   1.0225 &    & 1.938E+00 & 3.024E+08 & 9.807E+12 & 5.791E+16 & 5.729E+19
               & 3.657E+04 & 2.236E+09 & 1.522E+14 & 4.885E+18 & 3.798E+22 \\
   1.0226 & P5 & 2.724E+00 & 3.070E+08 & 3.247E+13 & 2.873E+18 & 3.918E+22
               & 3.658E+04 & 2.237E+09 & 1.949E+14 & 3.395E+19 & 2.051E+24 \\
   1.0281 &    & 2.695E+00 & 2.989E+08 & 3.128E+13 & 2.739E+18 & 3.698E+22
               & 3.600E+04 & 2.178E+09 & 1.878E+14 & 3.235E+19 & 1.934E+24 \\
   1.0282 & P4 & 2.890E+00 & 3.167E+08 & 4.031E+13 & 5.077E+18 & 7.818E+22
               & 3.604E+04 & 2.291E+09 & 1.969E+14 & 3.244E+19 & 1.932E+24 \\
\end{longtable}

%% file: Z117_TotIcc.tex
\renewcommand{\arraystretch}{1.0}


\setlength{\LTleft}{0pt}
\setlength{\LTright}{0pt}


\setlength{\tabcolsep}{0.5\tabcolsep}

\renewcommand{\arraystretch}{1.0}

\footnotesize 

\begin{longtable}{@{\extracolsep\fill}rlcccccccccc}
\caption[Total conversion coefficients for $Z=117$]{Total conversion coefficients for $Z=117$ (Us--$292$, Element $117$).
  See page \pageref{sec:TotIcc} for Explanation of Tables.
  \label{tab:Z117_TotIcc}}
E$_{\gamma}$ & Shell & E1 & M1 & E2 & M2 & E3 & M3 & E4 & M4 & E5 & M5 \\
\hline \\
\endfirsthead
\caption[]{Continued.}
E$_{\gamma}$ & Shell & E1 & M1 & E2 & M2 & E3 & M3 & E4 & M4 & E5 & M5 \\
\hline \\
\endhead
   1.0052 & Q3 & 4.064E-02 & 1.414E+08 & 3.155E+12 & 1.367E+16 & 1.013E+19
               & 3.101E+01 & 6.576E+08 & 1.918E+14 & 7.684E+18 & 6.420E+22 \\
   1.0124 & Q2 & 1.055E+00 & 4.017E+08 & 1.344E+13 & 8.218E+16 & 8.443E+19
               & 6.781E+03 & 6.932E+08 & 1.827E+14 & 7.210E+18 & 5.945E+22 \\
   1.0236 &    & 1.006E+00 & 3.803E+08 & 1.246E+13 & 7.463E+16 & 7.555E+19
               & 6.562E+03 & 6.562E+08 & 1.692E+14 & 6.535E+18 & 5.281E+22 \\
   1.0237 & Q1 & 2.285E+00 & 3.877E+08 & 1.250E+13 & 7.462E+16 & 7.549E+19
               & 4.485E+04 & 2.756E+09 & 2.024E+14 & 6.675E+18 & 5.288E+22 \\
   1.0274 &    & 2.251E+00 & 3.808E+08 & 1.219E+13 & 7.230E+16 & 7.279E+19
               & 4.436E+04 & 2.707E+09 & 1.974E+14 & 6.463E+18 & 5.087E+22 \\
   1.0275 & P5 & 3.056E+00 & 3.858E+08 & 3.831E+13 & 3.274E+18 & 4.480E+22
               & 4.437E+04 & 2.707E+09 & 2.473E+14 & 4.035E+19 & 2.406E+24 \\
   1.0336 &    & 3.018E+00 & 3.747E+08 & 3.676E+13 & 3.105E+18 & 4.203E+22
               & 4.360E+04 & 2.629E+09 & 2.373E+14 & 3.827E+19 & 2.255E+24 \\
   1.0337 & P4 & 3.207E+00 & 3.952E+08 & 4.687E+13 & 5.752E+18 & 8.912E+22
               & 4.364E+04 & 2.763E+09 & 2.479E+14 & 3.836E+19 & 2.253E+24 \\
\end{longtable}

%% file: Z118_TotIcc.tex
\renewcommand{\arraystretch}{1.0}


\setlength{\LTleft}{0pt}
\setlength{\LTright}{0pt}


\setlength{\tabcolsep}{0.5\tabcolsep}

\renewcommand{\arraystretch}{1.0}

\footnotesize 

\begin{longtable}{@{\extracolsep\fill}rlcccccccccc}
\caption[Total conversion coefficients for $Z=118$]{Total conversion coefficients for $Z=118$ (Uo--$294$, Element $118$).
  See page \pageref{sec:TotIcc} for Explanation of Tables.
  \label{tab:Z118_TotIcc}}
E$_{\gamma}$ & Shell & E1 & M1 & E2 & M2 & E3 & M3 & E4 & M4 & E5 & M5 \\
\hline \\
\endfirsthead
\caption[]{Continued.}
E$_{\gamma}$ & Shell & E1 & M1 & E2 & M2 & E3 & M3 & E4 & M4 & E5 & M5 \\
\hline \\
\endhead
   1.0060 & Q3 & 5.305E-02 & 1.813E+08 & 3.974E+12 & 1.730E+16 & 1.309E+19
               & 3.812E+01 & 8.539E+08 & 2.498E+14 & 1.011E+19 & 8.595E+22 \\
   1.0149 & Q2 & 1.262E+00 & 4.990E+08 & 1.658E+13 & 1.023E+17 & 1.074E+20
               & 8.785E+03 & 8.908E+08 & 2.352E+14 & 9.348E+18 & 7.817E+22 \\
   1.0271 &    & 1.199E+00 & 4.702E+08 & 1.527E+13 & 9.215E+16 & 9.515E+19
               & 8.477E+03 & 8.393E+08 & 2.164E+14 & 8.402E+18 & 6.874E+22 \\
   1.0272 & Q1 & 2.633E+00 & 4.791E+08 & 1.532E+13 & 9.214E+16 & 9.507E+19
               & 5.398E+04 & 3.306E+09 & 2.555E+14 & 8.568E+18 & 6.881E+22 \\
   1.0323 &    & 2.580E+00 & 4.675E+08 & 1.480E+13 & 8.823E+16 & 9.043E+19
               & 5.319E+04 & 3.225E+09 & 2.468E+14 & 8.197E+18 & 6.524E+22 \\
   1.0324 & P5 & 3.398E+00 & 4.729E+08 & 4.470E+13 & 3.704E+18 & 5.083E+22
               & 5.320E+04 & 3.226E+09 & 3.047E+14 & 4.742E+19 & 2.799E+24 \\
   1.0393 &    & 3.349E+00 & 4.575E+08 & 4.267E+13 & 3.490E+18 & 4.731E+22
               & 5.215E+04 & 3.120E+09 & 2.908E+14 & 4.467E+19 & 2.603E+24 \\
   1.0394 & P4 & 3.531E+00 & 4.812E+08 & 5.390E+13 & 6.461E+18 & 1.007E+23
               & 5.219E+04 & 3.278E+09 & 3.031E+14 & 4.478E+19 & 2.600E+24 \\
\end{longtable}

%% file: Z119_TotIcc.tex
\renewcommand{\arraystretch}{1.0}


\setlength{\LTleft}{0pt}
\setlength{\LTright}{0pt}


\setlength{\tabcolsep}{0.5\tabcolsep}

\renewcommand{\arraystretch}{1.0}

\footnotesize 

\begin{longtable}{@{\extracolsep\fill}rlcccccccccc}
\caption[Total conversion coefficients for $Z=119$]{Total conversion coefficients for $Z=119$ (A$=296$, Element $119$).
  See page \pageref{sec:TotIcc} for Explanation of Tables.
  \label{tab:Z119_TotIcc}}
E$_{\gamma}$ & Shell & E1 & M1 & E2 & M2 & E3 & M3 & E4 & M4 & E5 & M5 \\
\hline \\
\endfirsthead
\caption[]{Continued.}
E$_{\gamma}$ & Shell & E1 & M1 & E2 & M2 & E3 & M3 & E4 & M4 & E5 & M5 \\
\hline \\
\endhead
   1.0036 & R1 & 1.511E-01 & 1.058E+06 & 8.046E+09 & 9.176E+12 & 1.920E+15
               & 4.952E+03 & 2.800E+08 & 4.692E+12 & 2.197E+16 & 2.013E+19 \\
   1.0084 & Q3 & 2.192E-01 & 2.428E+08 & 5.192E+12 & 2.261E+16 & 1.746E+19
               & 4.931E+03 & 1.425E+09 & 3.416E+14 & 1.375E+19 & 1.184E+23 \\
   1.0196 &    & 2.067E-01 & 2.298E+08 & 4.809E+12 & 2.052E+16 & 1.561E+19
               & 4.771E+03 & 1.349E+09 & 3.163E+14 & 1.246E+19 & 1.051E+23 \\
   1.0197 & Q2 & 1.657E+00 & 6.310E+08 & 2.059E+13 & 1.274E+17 & 1.359E+20
               & 1.621E+04 & 1.441E+09 & 3.164E+14 & 1.245E+19 & 1.050E+23 \\
   1.0330 &    & 1.569E+00 & 5.916E+08 & 1.883E+13 & 1.137E+17 & 1.192E+20
               & 1.560E+04 & 1.351E+09 & 2.891E+14 & 1.109E+19 & 9.133E+22 \\
   1.0331 & Q1 & 3.201E+00 & 6.024E+08 & 1.889E+13 & 1.137E+17 & 1.191E+20
               & 7.066E+04 & 4.287E+09 & 3.354E+14 & 1.129E+19 & 9.142E+22 \\
   1.0398 &    & 3.118E+00 & 5.834E+08 & 1.807E+13 & 1.074E+17 & 1.116E+20
               & 6.931E+04 & 4.152E+09 & 3.206E+14 & 1.065E+19 & 8.527E+22 \\
   1.0399 & P5 & 3.958E+00 & 5.892E+08 & 5.165E+13 & 4.093E+18 & 5.600E+22
               & 6.932E+04 & 4.152E+09 & 3.863E+14 & 5.497E+19 & 3.167E+24 \\
\end{longtable}

%% file: Z120_TotIcc.tex
\renewcommand{\arraystretch}{1.0}


\setlength{\LTleft}{0pt}
\setlength{\LTright}{0pt}


\setlength{\tabcolsep}{0.5\tabcolsep}

\renewcommand{\arraystretch}{1.0}

\footnotesize 

\begin{longtable}{@{\extracolsep\fill}rlcccccccccc}
\caption[Total conversion coefficients for $Z=120$]{Total conversion coefficients for $Z=120$ (A$=301$, Element $120$).
  See page \pageref{sec:TotIcc} for Explanation of Tables.
  \label{tab:Z120_TotIcc}}
E$_{\gamma}$ & Shell & E1 & M1 & E2 & M2 & E3 & M3 & E4 & M4 & E5 & M5 \\
\hline \\
\endfirsthead
\caption[]{Continued.}
E$_{\gamma}$ & Shell & E1 & M1 & E2 & M2 & E3 & M3 & E4 & M4 & E5 & M5 \\
\hline \\
\endhead
   1.0042 & R1 & 2.183E-01 & 1.648E+06 & 1.267E+10 & 1.480E+13 & 3.161E+15
               & 7.599E+03 & 4.271E+08 & 7.204E+12 & 3.428E+16 & 3.228E+19 \\
   1.0113 & Q3 & 3.019E-01 & 3.066E+08 & 6.397E+12 & 2.787E+16 & 2.195E+19
               & 7.500E+03 & 1.880E+09 & 4.360E+14 & 1.761E+19 & 1.537E+23 \\
   1.0250 &    & 2.820E-01 & 2.866E+08 & 5.828E+12 & 2.476E+16 & 1.914E+19
               & 7.205E+03 & 1.758E+09 & 3.969E+14 & 1.562E+19 & 1.329E+23 \\
   1.0251 & Q2 & 1.984E+00 & 7.722E+08 & 2.485E+13 & 1.544E+17 & 1.677E+20
               & 2.192E+04 & 1.872E+09 & 3.971E+14 & 1.561E+19 & 1.328E+23 \\
   1.0395 &    & 1.873E+00 & 7.204E+08 & 2.256E+13 & 1.366E+17 & 1.456E+20
               & 2.103E+04 & 1.746E+09 & 3.603E+14 & 1.378E+19 & 1.143E+23 \\
   1.0396 & Q1 & 3.713E+00 & 7.334E+08 & 2.265E+13 & 1.366E+17 & 1.454E+20
               & 8.705E+04 & 5.206E+09 & 4.146E+14 & 1.401E+19 & 1.144E+23 \\
   1.0478 &    & 3.599E+00 & 7.053E+08 & 2.145E+13 & 1.275E+17 & 1.343E+20
               & 8.503E+04 & 5.007E+09 & 3.925E+14 & 1.306E+19 & 1.051E+23 \\
   1.0479 & P5 & 4.457E+00 & 7.115E+08 & 5.890E+13 & 4.480E+18 & 6.111E+22
               & 8.503E+04 & 5.006E+09 & 4.664E+14 & 6.264E+19 & 3.542E+24 \\
\end{longtable}

%% file: Z121_TotIcc.tex
\renewcommand{\arraystretch}{1.0}


\setlength{\LTleft}{0pt}
\setlength{\LTright}{0pt}


\setlength{\tabcolsep}{0.5\tabcolsep}

\renewcommand{\arraystretch}{1.0}

\footnotesize 

\begin{longtable}{@{\extracolsep\fill}rlcccccccccc}
\caption[Total conversion coefficients for $Z=121$]{Total conversion coefficients for $Z=121$ (A$=300$, Element $121$).
  See page \pageref{sec:TotIcc} for Explanation of Tables.
  \label{tab:Z121_TotIcc}}
E$_{\gamma}$ & Shell & E1 & M1 & E2 & M2 & E3 & M3 & E4 & M4 & E5 & M5 \\
\hline \\
\endfirsthead
\caption[]{Continued.}
E$_{\gamma}$ & Shell & E1 & M1 & E2 & M2 & E3 & M3 & E4 & M4 & E5 & M5 \\
\hline \\
\endhead
   1.0030 & R2 & 1.528E-01 & 4.638E+07 & 1.907E+12 & 1.375E+16 & 1.680E+19
               & 1.418E+03 & 1.124E+07 & 4.591E+10 & 4.137E+13 & 7.759E+15 \\
   1.0054 & R1 & 7.529E-01 & 5.075E+07 & 1.914E+12 & 1.351E+16 & 1.640E+19
               & 2.381E+04 & 1.258E+09 & 2.118E+13 & 1.021E+17 & 9.858E+19 \\
   1.0150 &    & 7.262E-01 & 4.841E+07 & 1.792E+12 & 1.243E+16 & 1.489E+19
               & 2.314E+04 & 1.200E+09 & 1.983E+13 & 9.389E+16 & 8.949E+19 \\
   1.0151 & Q3 & 8.373E-01 & 4.306E+08 & 9.579E+12 & 4.628E+16 & 4.205E+19
               & 2.321E+04 & 3.057E+09 & 5.616E+14 & 2.235E+19 & 1.966E+23 \\
   1.0313 &    & 7.817E-01 & 3.979E+08 & 8.583E+12 & 4.028E+16 & 3.580E+19
               & 2.213E+04 & 2.825E+09 & 5.029E+14 & 1.940E+19 & 1.658E+23 \\
   1.0314 & Q2 & 2.759E+00 & 9.809E+08 & 3.127E+13 & 1.953E+17 & 2.163E+20
               & 4.098E+04 & 2.965E+09 & 5.031E+14 & 1.939E+19 & 1.656E+23 \\
   1.0470 &    & 2.597E+00 & 9.103E+08 & 2.818E+13 & 1.712E+17 & 1.857E+20
               & 3.918E+04 & 2.751E+09 & 4.531E+14 & 1.695E+19 & 1.409E+23 \\
   1.0471 & Q1 & 4.643E+00 & 9.258E+08 & 2.828E+13 & 1.712E+17 & 1.855E+20
               & 1.177E+05 & 6.783E+09 & 5.158E+14 & 1.722E+19 & 1.410E+23 \\
\end{longtable}

%% file: Z122_TotIcc.tex
\renewcommand{\arraystretch}{1.0}


\setlength{\LTleft}{0pt}
\setlength{\LTright}{0pt}


\setlength{\tabcolsep}{0.5\tabcolsep}

\renewcommand{\arraystretch}{1.0}

\footnotesize 

\begin{longtable}{@{\extracolsep\fill}rlcccccccccc}
\caption[Total conversion coefficients for $Z=122$]{Total conversion coefficients for $Z=122$ (A$=305$, Element $122$).
  See page \pageref{sec:TotIcc} for Explanation of Tables.
  \label{tab:Z122_TotIcc}}
E$_{\gamma}$ & Shell & E1 & M1 & E2 & M2 & E3 & M3 & E4 & M4 & E5 & M5 \\
\hline \\
\endfirsthead
\caption[]{Continued.}
E$_{\gamma}$ & Shell & E1 & M1 & E2 & M2 & E3 & M3 & E4 & M4 & E5 & M5 \\
\hline \\
\endhead
   1.0033 & R2 & 2.295E-01 & 7.492E+07 & 4.665E+12 & 5.534E+17 & 1.110E+22
               & 2.296E+03 & 4.522E+07 & 2.274E+12 & 2.993E+16 & 6.052E+19 \\
   1.0062 & R1 & 5.998E-01 & 7.714E+07 & 4.598E+12 & 5.393E+17 & 1.076E+22
               & 1.703E+04 & 8.609E+08 & 1.614E+13 & 9.732E+16 & 1.263E+20 \\
   1.0176 &    & 5.758E-01 & 7.294E+07 & 4.252E+12 & 4.877E+17 & 9.529E+21
               & 1.647E+04 & 8.140E+08 & 1.493E+13 & 8.815E+16 & 1.126E+20 \\
   1.0177 & Q3 & 7.098E-01 & 5.358E+08 & 1.344E+13 & 5.272E+17 & 9.552E+21
               & 1.655E+04 & 3.079E+09 & 6.763E+14 & 2.740E+19 & 2.445E+23 \\
   1.0366 &    & 6.553E-01 & 4.888E+08 & 1.183E+13 & 4.474E+17 & 7.832E+21
               & 1.566E+04 & 2.810E+09 & 5.949E+14 & 2.325E+19 & 2.005E+23 \\
   1.0367 & Q2 & 2.899E+00 & 1.173E+09 & 3.834E+13 & 6.292E+17 & 8.040E+21
               & 3.920E+04 & 2.978E+09 & 5.951E+14 & 2.323E+19 & 2.003E+23 \\
   1.0533 &    & 2.722E+00 & 1.084E+09 & 3.434E+13 & 5.465E+17 & 6.776E+21
               & 3.738E+04 & 2.751E+09 & 5.327E+14 & 2.016E+19 & 1.688E+23 \\
   1.0534 & Q1 & 4.981E+00 & 1.102E+09 & 3.446E+13 & 5.461E+17 & 6.769E+21
               & 1.297E+05 & 7.411E+09 & 6.049E+14 & 2.047E+19 & 1.689E+23 \\
\end{longtable}

%% file: Z123_TotIcc.tex
\renewcommand{\arraystretch}{1.0}


\setlength{\LTleft}{0pt}
\setlength{\LTright}{0pt}


\setlength{\tabcolsep}{0.5\tabcolsep}

\renewcommand{\arraystretch}{1.0}

\footnotesize 

\begin{longtable}{@{\extracolsep\fill}rlcccccccccc}
\caption[Total conversion coefficients for $Z=123$]{Total conversion coefficients for $Z=123$ (A$=304$, Element $123$).
  See page \pageref{sec:TotIcc} for Explanation of Tables.
  \label{tab:Z123_TotIcc}}
E$_{\gamma}$ & Shell & E1 & M1 & E2 & M2 & E3 & M3 & E4 & M4 & E5 & M5 \\
\hline \\
\endfirsthead
\caption[]{Continued.}
E$_{\gamma}$ & Shell & E1 & M1 & E2 & M2 & E3 & M3 & E4 & M4 & E5 & M5 \\
\hline \\
\endhead
   1.0034 & R2 & 1.902E-01 & 6.305E+07 & 2.616E+12 & 1.941E+16 & 2.486E+19
               & 2.145E+03 & 1.598E+07 & 6.610E+10 & 6.206E+13 & 1.189E+16 \\
   1.0047 & P6 & 3.707E+00 & 6.291E+07 & 2.700E+12 & 3.696E+16 & 2.755E+22
               & 2.141E+03 & 1.596E+07 & 4.778E+11 & 1.046E+17 & 3.076E+21 \\
   1.0063 &    & 3.704E+00 & 6.242E+07 & 2.671E+12 & 3.645E+16 & 2.707E+22
               & 2.131E+03 & 1.583E+07 & 4.725E+11 & 1.031E+17 & 3.024E+21 \\
   1.0064 & R1 & 4.062E+00 & 6.581E+07 & 2.696E+12 & 3.645E+16 & 2.704E+22
               & 1.728E+04 & 8.472E+08 & 1.472E+13 & 1.738E+17 & 3.092E+21 \\
   1.0183 &    & 4.029E+00 & 6.207E+07 & 2.485E+12 & 3.283E+16 & 2.378E+22
               & 1.668E+04 & 7.991E+08 & 1.357E+13 & 1.565E+17 & 2.723E+21 \\
   1.0184 & Q3 & 4.175E+00 & 5.607E+08 & 1.215E+13 & 7.486E+16 & 2.379E+22
               & 1.676E+04 & 3.260E+09 & 7.335E+14 & 3.014E+19 & 2.750E+23 \\
   1.0393 &    & 4.094E+00 & 5.067E+08 & 1.056E+13 & 6.257E+16 & 1.904E+22
               & 1.577E+04 & 2.946E+09 & 6.366E+14 & 2.514E+19 & 2.209E+23 \\
   1.0394 & Q2 & 6.576E+00 & 1.294E+09 & 4.100E+13 & 2.737E+17 & 1.928E+22
               & 4.451E+04 & 3.144E+09 & 6.370E+14 & 2.512E+19 & 2.207E+23 \\
\end{longtable}

%% file: Z124_TotIcc.tex
\renewcommand{\arraystretch}{1.0}


\setlength{\LTleft}{0pt}
\setlength{\LTright}{0pt}


\setlength{\tabcolsep}{0.5\tabcolsep}

\renewcommand{\arraystretch}{1.0}

\footnotesize 

\begin{longtable}{@{\extracolsep\fill}rlcccccccccc}
\caption[Total conversion coefficients for $Z=124$]{Total conversion coefficients for $Z=124$ (A$=309$, Element $124$).
  See page \pageref{sec:TotIcc} for Explanation of Tables.
  \label{tab:Z124_TotIcc}}
E$_{\gamma}$ & Shell & E1 & M1 & E2 & M2 & E3 & M3 & E4 & M4 & E5 & M5 \\
\hline \\
\endfirsthead
\caption[]{Continued.}
E$_{\gamma}$ & Shell & E1 & M1 & E2 & M2 & E3 & M3 & E4 & M4 & E5 & M5 \\
\hline \\
\endhead
   1.0033 & R2 & 2.703E-01 & 9.410E+07 & 3.925E+12 & 2.958E+16 & 3.886E+19
               & 3.372E+03 & 2.440E+07 & 1.017E+11 & 9.762E+13 & 1.908E+16 \\
   1.0034 & P6 & 4.581E+00 & 9.440E+07 & 4.075E+12 & 5.339E+16 & 3.883E+22
               & 3.377E+03 & 2.450E+07 & 6.941E+11 & 1.497E+17 & 4.473E+21 \\
   1.0060 &    & 4.576E+00 & 9.319E+07 & 4.003E+12 & 5.217E+16 & 3.775E+22
               & 3.351E+03 & 2.419E+07 & 6.817E+11 & 1.463E+17 & 4.350E+21 \\
   1.0061 & R1 & 5.023E+00 & 9.778E+07 & 4.037E+12 & 5.218E+16 & 3.771E+22
               & 2.353E+04 & 1.128E+09 & 1.972E+13 & 2.423E+17 & 4.445E+21 \\
   1.0173 &    & 4.985E+00 & 9.255E+07 & 3.739E+12 & 4.729E+16 & 3.340E+22
               & 2.277E+04 & 1.067E+09 & 1.826E+13 & 2.196E+17 & 3.944E+21 \\
   1.0174 & Q3 & 5.159E+00 & 7.015E+08 & 1.531E+13 & 9.784E+16 & 3.341E+22
               & 2.286E+04 & 4.087E+09 & 9.096E+14 & 3.778E+19 & 3.517E+23 \\
   1.0396 &    & 5.055E+00 & 6.299E+08 & 1.318E+13 & 8.086E+16 & 2.638E+22
               & 2.143E+04 & 3.670E+09 & 7.825E+14 & 3.116E+19 & 2.786E+23 \\
   1.0397 & Q2 & 7.907E+00 & 1.580E+09 & 5.008E+13 & 3.403E+17 & 2.667E+22
               & 5.799E+04 & 3.914E+09 & 7.829E+14 & 3.113E+19 & 2.784E+23 \\
\end{longtable}

%% file: Z125_TotIcc.tex
\renewcommand{\arraystretch}{1.0}


\setlength{\LTleft}{0pt}
\setlength{\LTright}{0pt}


\setlength{\tabcolsep}{0.5\tabcolsep}

\renewcommand{\arraystretch}{1.0}

\footnotesize 

\begin{longtable}{@{\extracolsep\fill}rlcccccccccc}
\caption[Total conversion coefficients for $Z=125$]{Total conversion coefficients for $Z=125$ (A$=308$, Element $125$).
  See page \pageref{sec:TotIcc} for Explanation of Tables.
  \label{tab:Z125_TotIcc}}
E$_{\gamma}$ & Shell & E1 & M1 & E2 & M2 & E3 & M3 & E4 & M4 & E5 & M5 \\
\hline \\
\endfirsthead
\caption[]{Continued.}
E$_{\gamma}$ & Shell & E1 & M1 & E2 & M2 & E3 & M3 & E4 & M4 & E5 & M5 \\
\hline \\
\endhead
   1.0068 & R2 & 2.857E-01 & 1.036E+08 & 4.304E+12 & 3.260E+16 & 4.321E+19
               & 3.952E+03 & 2.745E+07 & 1.140E+11 & 1.100E+14 & 2.110E+16 \\
   1.0086 & O8 & 1.002E+02 & 1.028E+08 & 4.252E+12 & 3.248E+16 & 6.562E+19
               & 3.935E+03 & 2.722E+07 & 1.128E+11 & 1.274E+15 & 6.210E+20 \\
   1.0088 &    & 1.001E+02 & 1.027E+08 & 4.246E+12 & 3.243E+16 & 6.548E+19
               & 3.932E+03 & 2.719E+07 & 1.127E+11 & 1.272E+15 & 6.196E+20 \\
   1.0089 & P6 & 1.041E+02 & 1.030E+08 & 4.407E+12 & 5.629E+16 & 4.045E+22
               & 3.937E+03 & 2.729E+07 & 7.609E+11 & 1.617E+17 & 5.414E+21 \\
   1.0105 &    & 1.035E+02 & 1.022E+08 & 4.359E+12 & 5.551E+16 & 3.975E+22
               & 3.919E+03 & 2.707E+07 & 7.526E+11 & 1.594E+17 & 5.321E+21 \\
   1.0106 & R1 & 1.040E+02 & 1.073E+08 & 4.397E+12 & 5.551E+16 & 3.971E+22
               & 2.639E+04 & 1.235E+09 & 2.151E+13 & 2.644E+17 & 5.426E+21 \\
   1.0230 &    & 9.978E+01 & 1.010E+08 & 4.041E+12 & 4.982E+16 & 3.474E+22
               & 2.544E+04 & 1.162E+09 & 1.976E+13 & 2.373E+17 & 4.755E+21 \\
   1.0231 & Q3 & 9.993E+01 & 7.469E+08 & 1.588E+13 & 1.009E+17 & 3.475E+22
               & 2.553E+04 & 4.383E+09 & 9.636E+14 & 3.992E+19 & 3.735E+23 \\
\end{longtable}

%% file: Z126_TotIcc.tex
\renewcommand{\arraystretch}{1.0}


\setlength{\LTleft}{0pt}
\setlength{\LTright}{0pt}


\setlength{\tabcolsep}{0.5\tabcolsep}

\renewcommand{\arraystretch}{1.0}

\footnotesize 

\begin{longtable}{@{\extracolsep\fill}rlcccccccccc}
\caption[Total conversion coefficients for $Z=126$]{Total conversion coefficients for $Z=126$ (A$=310$, Element $126$).
  See page \pageref{sec:TotIcc} for Explanation of Tables.
  \label{tab:Z126_TotIcc}}
E$_{\gamma}$ & Shell & E1 & M1 & E2 & M2 & E3 & M3 & E4 & M4 & E5 & M5 \\
\hline \\
\endfirsthead
\caption[]{Continued.}
E$_{\gamma}$ & Shell & E1 & M1 & E2 & M2 & E3 & M3 & E4 & M4 & E5 & M5 \\
\hline \\
\endhead
   1.0039 & R2 & 2.365E-01 & 9.072E+07 & 3.800E+12 & 2.929E+16 & 3.968E+19
               & 3.639E+03 & 2.455E+07 & 1.029E+11 & 1.011E+14 & 1.922E+16 \\
   1.0059 & P6 & 3.750E+00 & 9.016E+07 & 3.918E+12 & 5.235E+16 & 4.211E+22
               & 3.623E+03 & 2.442E+07 & 7.879E+11 & 1.694E+17 & 5.141E+21 \\
   1.0060 &    & 3.750E+00 & 9.012E+07 & 3.915E+12 & 5.230E+16 & 4.206E+22
               & 3.622E+03 & 2.441E+07 & 7.873E+11 & 1.692E+17 & 5.136E+21 \\
   1.0061 & O8 & 1.098E+02 & 9.022E+07 & 3.914E+12 & 5.273E+16 & 4.204E+22
               & 3.625E+03 & 2.440E+07 & 7.870E+11 & 1.705E+17 & 5.893E+21 \\
   1.0073 &    & 1.094E+02 & 8.969E+07 & 3.882E+12 & 5.217E+16 & 4.150E+22
               & 3.612E+03 & 2.426E+07 & 7.805E+11 & 1.687E+17 & 5.818E+21 \\
   1.0074 & R1 & 1.098E+02 & 9.489E+07 & 3.921E+12 & 5.218E+16 & 4.145E+22
               & 2.587E+04 & 1.222E+09 & 2.160E+13 & 2.759E+17 & 5.926E+21 \\
   1.0201 &    & 1.052E+02 & 8.915E+07 & 3.595E+12 & 4.669E+16 & 3.614E+22
               & 2.492E+04 & 1.148E+09 & 1.980E+13 & 2.468E+17 & 5.175E+21 \\
   1.0202 & Q3 & 1.054E+02 & 7.658E+08 & 1.575E+13 & 9.957E+16 & 3.615E+22
               & 2.501E+04 & 4.534E+09 & 1.023E+15 & 4.305E+19 & 4.117E+23 \\
\end{longtable}

%% file: TableReferences.tex
\begin{theDTbibliography}{1956He83}

\bibitem{AtomicWeights}
   \emph{"Atomic Weights of the Elements 2007"},
   IUPAC Commission on Atomic Weights and Isotopic Abundances,
    http://www.chem.qmul.ac.uk/iupac/AtWt/,
    prepared by G.P. Moss,
    based on: Pure Appl. Chem. \textbf{78} (2006) 2051.
\bibitem{2002Ba85}
    I.M.~Band, M.B.~Trzhaskovskaya, C.W.~Nestor, Jr.,
    P.O.~Tikkanen, S.~Raman,
    At. Data and Nucl. Data Tables {\textbf 81} (2002) 1.
\bibitem{2006Ne11}
    V.I.~Nefedov, M.B.~Trzhaskovskaya, V.G.~Yarzhemskii,
    Doklady Physical Chemistry \textbf{408} (2006) 149;
    Dok. Aka. Nauk \textbf{408} (2006) 488.
\bibitem{1977Ca31}
    T.A. Carlson, C.W. Nestor, Jr,
    At. Data and Nucl. Data Tables \textbf{19} (1977) 153.
\bibitem{1977Fr21}
    B. Fricke, G. Soff,
    At. Data and Nucl. Data Tables \textbf{19} (1977) 83.
\bibitem{2008Ki07}
   T. Kib\'{e}di, T.W. Burrows, M.B. Trzhaskovskaya, P.M. Davidson, C.W. Nestor, Jr.,
   Nucl. Instr.  Meth. in Phys. Res. A  {\textbf 589} (2008) 202.
\bibitem{2001Landau}
    A. Landau, E. Eliav, Y. Ishikawa, U. Kaldor,
    J. Chem. Phys. \textbf{114} (2001) 2977.
\bibitem{1998Eliav}
    E. Eliav, U. Kaldor, Y. Ishikawa,
    Mol. Phys. \textbf{94} (1998) 181.
\bibitem{1975Fricke}
    B. Fricke,
    Struct. Bond. \textbf{21} (1975) 89.
\bibitem{2008Pershina}
    V. Pershina, A. Borschevsky, E. Eliav, U. Kaldor,
    J. Chem. Phys. \textbf{129} (2008) 144106.
\end{theDTbibliography}